\documentstyle[aps,prd,epsfig,eqsecnum]{revtex}

\newcommand{\talpha}{\tilde{\alpha}}
\newcommand{\tmu}{\tilde{\mu}}

\newcommand{\cf}{{\it cf.}}
\newcommand{\e}{{\rm e}}
\newcommand{\bb}{\bbox}
\newcommand{\nn}{\nonumber \\}
\newcommand{\be}{\begin{eqnarray}}
\newcommand{\ba}{\begin{array}}
\newcommand{\ea}{\end{array}}
\newcommand{\ee}{\end{eqnarray}}

\newcommand{\xp}{x^\prime}
\newcommand{\xpp}{x^{\prime\prime}}

\newcommand{\dy}{(dy)}

\newcommand{\dzi}{(dz_1)}
\newcommand{\dzii}{(dz_2)}
\newcommand{\dxi}{(dx_1)}
\newcommand{\dxii}{(dx_2)}
\newcommand{\dyi}{(dy_1)}
\newcommand{\dyii}{(dy_2)}
\newcommand{\pp}{p^{\prime} }

\newcommand{\ppi}{p^{\prime}_{1} }
\newcommand{\ppii}{p^{\prime}_{2} }
\newcommand{\dx}{(dx)}
\newcommand{\dxp}{(dx^{\prime})}
\newcommand{\dxpp}{(dx^{\prime\prime})}

\newcommand{\dslash}{\gamma\partial}

\newcommand{\Aslash}{\gamma A}
\newcommand{\Bslash}{\gamma B}
\newcommand{\bAslash}{\gamma \bar{A}}
\newcommand{\bBslash}{\gamma \bar{B}}

\newcommand{\cq}{{{\mathsf{q}}_1\cdot{\mathsf{q}}_2}}
\newcommand{\wq}{{{\mathsf{q}}_1\bbox{\times}{\mathsf{q}}_2}}
\newfont{\fib}{cmfi10 at 10pt}
\newcommand{\ps}{\mbox{\fib{s}}} 
\newcommand{\pt}{\mbox{\fib{t}}}

\newcommand{\Tr}{{\rm Tr}}

\newcommand{\A}{{\cal A}}

\newcommand{\bA}{{\bar{A}}}

\newcommand{\tD}{{\tilde{D}}}
\newcommand{\tA}{\tilde{A}}
\newcommand{\tB}{\tilde{B}}
\newcommand{\D}{{\cal D}}
\newcommand{\N}{{\cal N}}
\newcommand{\J}{{\cal J}}
\newcommand{\K}{{\cal K}}

\newcommand{\eg}{{\it e.g.}\ }

\newcommand{\fg}{\hspace{.1cm}^{\ast}F}
\newcommand{\jg}{\hskip .05cm ^{\ast}\hskip -.07cm j}
\newcommand{\Jg}{\hskip .05cm ^{\ast}\hskip -.07cm J}
\newcommand{\Gg}{\hspace{.1cm}^{\ast}G}

\newcommand{\dphi}{\left[{\rm d}\phi\right]}
\newcommand{\dphip}{\left[{\rm d}\phi^\prime\right]}

\begin{document}

\draft
\preprint{}

\title{\hfill{\rm OKHEP--99--02}\\
Dual Quantum Electrodynamics:\\
Dyon-Dyon and Charge-Monopole  Scattering in a High-Energy
Approximation}
\author{Leonard Gamberg\thanks{Electronic address: gamberg@mail.nhn.ou.edu}
and Kimball A. Milton\thanks{Electronic address: milton@mail.nhn.ou.edu}}
\address{Department of Physics and Astronomy, University of Oklahoma, Norman, 
OK 73019 USA}

\date{\today}
\maketitle

\begin{abstract}
We develop the quantum field theory of electron-point
magnetic monopole interactions and more generally, dyon-dyon interactions,
 based on the original
string-dependent ``nonlocal'' action of Dirac and Schwinger.
We demonstrate that a viable nonperturbative quantum
field theoretic formulation
can be constructed that results in a string {\em independent}
cross section for monopole-electron and dyon-dyon scattering. 
Such calculations can be done only by using nonperturbative
approximations such as the eikonal and not by some
mutilation of lowest-order perturbation theory.
\end{abstract}
\pacs{PACS number(s): 14.80.Hv,11.80.Fv,12.20.Ds,11.30.Fs}

\section{Introduction}
The subject of magnetic charge in quantum mechanics has been of great
interest since the work of Dirac~\cite{dir31}, who showed that electric
and magnetic charge could co-exist provided the quantization condition
(in rationalized units)
\be
\frac{eg}{4\pi}=\frac{N}{2} 
\hspace{.25cm},\hspace{.25cm} N \in\,  Z\, ,
\label{quan}
\ee
holds, where $e$ and $g$ are the electric and magnetic charges,
respectively, $Z$ being the integers.  In addition, the existence of
topological or  ``extended'' magnetic 
charge has been demonstrated in non-Abelian gauge 
theories~\cite{nonabel}, where they can have profound effects,
most notably in the infrared regime of the QCD 
vacuum.  For example, in the 
illustrative scenario of color confinement  proposed
by Mandelstam and 't Hooft~\cite{man76} it is conjectured 
that the QCD vacuum behaves as a {\em dual} type II superconductor.
Due to 
the condensation  of magnetic monopoles which emerge from the {\em Abelian} 
projection~\cite{gia98} of the SU(3) color gauge group,
the chromo-electric field acting between $q\bar q$ pairs 
is squeezed into dual Abrikosov flux tubes~\cite{ab57}.
Very recent studies suggest that these flux tubes may determine the
infrared properties of the gluon propagator \cite{GPZ}.
Also it has been suggested~\cite{rei98} that color monopoles 
are sufficient to trigger the spontaneous breakdown of chiral symmetry.

The monopoles most commonly considered today as being potentially observable 
are those associated 
with the grand-unification  symmetry breaking scale, monopoles having masses 
of order $10^{16}$ GeV; however, much lower symmetry-breaking scales 
could be relevant, giving rise to monopoles of mass of order 
10 TeV~\cite{preskill}.  [It is worthwhile mentioning that the
previously-believed prohibition ({\cf} Ref.~\cite{col81}) against
topological monopoles
in electroweak theory is apparently too  restrictive~\cite{cho97}, 
thus raising 
the possibility  of the existence of monopoles at the  electroweak-symmetry 
breaking scale.] Further, there is no reason 
why ``elementary'' Dirac monopoles should not exist; only 
experiment  can settle this question.

Since the early 1970s there have been many experimental searches for
magnetic monopoles, ranging from seeking cosmological bounds to 
studies of lunar samples~\cite{searches}. 
Although none of these searches ultimately
has yielded a positive signal, the arguments in favor of the existence
of magnetic charge, whether elementary or ``extended,'' remain as cogent
as ever.   However, the quantum field theory of elementary or pointlike 
magnetic charges remains poorly developed, particularly at the 
phenomenological level. In view of the necessity of 
establishing a reliable estimate for monopole production in accelerators, 
for example, in order to set bounds on monopole masses in terrestrial 
experiments~\cite{kal95},  it is important to put the theory on a 
firmer foundation.  (This is especially so for limits set through 
virtual-monopole processes \cite{ginzd0}.  For a critique of such limits 
see Ref.~\cite{gamberg}.) It is the purpose of this paper to initiate a 
complete study of monopole scattering and production in relativistic 
quantum electrodynamics extended to include point magnetic 
charges, ``dual QED.''  Given ongoing experiments it is important to 
obtain reliable calculations of scattering processes.

From the work of Dirac on the nonrelativistic
and the relativistic quantum mechanics 
of magnetic monopoles \cite{dir31,dir48}, and the subsequent 
work of Schwinger\cite{sch66,sch66r,sch68} and later Zwanziger\cite{zwa71}
on relativistic quantum field theories of magnetic monopoles, it is 
known that it is not possible to write down a theory of 
pointlike electric and magnetic currents interacting via the electromagnetic 
field described solely in terms of a local vector potential ${A}_{\mu}(x)$.  
Consistency of the Maxwell equations
\be
\partial^\nu F_{\mu\nu}=j_\mu \hspace{.5cm} {\mbox {\rm and}} \hspace{.5cm}
\partial^{\nu}\fg_{\mu\nu}=\jg_{\mu}\ , 
\label{max}
\ee
where 
$\fg_{\mu\nu}={1\over2}\epsilon_{\mu\nu\sigma\tau}F^{\sigma\tau}$,
which imply the dual conservation of electric and magnetic
currents, $j_\mu$ and $\jg_\mu$, respectively,
\be
\partial_\mu j^{\mu}=0 \hspace{.5cm} {\mbox {\rm and}} \hspace{.5cm}
\partial_\mu \jg^{\mu}=0\,  ,
\label{con}
\ee
necessitates the introduction of the Dirac string \cite{dir31,sch66} 
or a multi-valued potential \cite{wu75}. 
In this paper we will follow the string-dependent  formulation, where
the singular electromagnetic potential is consistent with
nonrelativistic quantum mechanics  provided the Dirac quantization
condition (\ref{quan}) holds, or, for particles (labeled by $a$, $b$, etc.) 
carrying both electric
and magnetic charge (dyons), the Schwinger generalization
\be
\frac{e_{a}g_{b}-e_{b}g_{a}}{4\pi}=
\Bigg\{
\begin{array}{l}
\frac{N}{2}\, , {\mbox{unsymmetric}}\\
\mbox{\small $N$}\, , \mbox{symmetric}
\end{array}
\Bigg\}, \quad \mbox{{\small $N$}} \in\, {\mbox{$Z$}}
\label{dyon}
\ee 
is invoked. (``Symmetric'' and ``unsymmetric'' refer to the presence or
absence of dual symmetry in the solutions of Maxwell's equations.)
 That is, consistency conditions, Eqs.~(\ref{quan}) and 
(\ref{dyon}), must necessarily be satisfied if physical observables 
are to be rendered independent of a gauge artifact, the Dirac string 
singularity.  {\em Formally}, rotational invariance of nonrelativistic quantum 
mechanics with monopoles follows directly from the Schr\"{o}dinger equation 
where re-orienting the string is equivalent to a gauge transformation on the 
wavefunction which in turn is well-defined given (\ref{quan}) or (\ref{dyon}).
In addition, in early work on the nonrelativistic  
electron-monopole scattering by Goldhaber~\cite{gold65},
 Schwinger {\it et al.} \cite{mil76}, and  
Milton and DeRaad \cite{mil78}, not surprisingly, it was found that
the resulting scattering 
cross section is gauge-string-independent if and only if
the quantization condition is satisfied.\footnote{Upon separation
of variables, the angular differential operator contains a contribution from 
the ``intrinsic'' field angular momentum carried by the  interaction of the 
monopole  with the photon field, which is naturally quantized
upon using  the Dirac-Schwinger condition  (\ref{dyon}); that is, 
there is an  additional effective magnetic quantum number
$m^{\prime}=\frac{e_ag_b-e_ag_b}{4\pi}$, a fact already implicit in the
classical analysis of Poincar\'e and Thomson \cite{history}.}

On the other hand, attempts to incorporate monopoles consistently into
relativistic quantum field theory have met with mixed success.  
Weinberg~\cite{wei65} and, somehwat thereafter, Rabl\cite{rab69}  demonstrated 
that electron--monopole scattering calculated in the
one-photon-exchange approximation was described by
 a string-dependent scattering 
amplitude.  A straightforward calculation begins, for example, with
the interaction given by Schwinger~\cite{sch68,sch75}
\be
W(j,\jg)&=&\int \dx \dxp \dxpp
\jg^\mu(x)\epsilon_{\mu\nu\sigma\tau}
\partial^{\nu} f^{\sigma}\left(x-x^{\prime}\right)
D_{+}\left(x^{\prime}-x^{\prime\prime}\right)j^{\tau}
\left(x^{\prime\prime}\right)\ .
\label{interact}
\ee
Here the electric and magnetic currents are
$j_\mu=e\bar\psi\gamma_\mu\psi$ and $\jg_\mu=g\bar\chi\gamma_\mu\chi$,
for example, for spin-1/2 particles.
The photon propagator is denoted by $D_+(x-x')$ and $f_\mu(x)$ is 
the Dirac string function which satisfies the differential equation 
\be
\partial_\mu f^{\mu}(x)=\delta(x)\, ,
\label{dfstng}
\ee
a formal solution of which is given by 
\be
f^{\mu}(x)=n^\mu\left(n\cdot\partial\right)^{-1}\delta(x)\,,
\label{stngsl}
\ee
where $n^\mu$ is an arbitrary vector.
Therefore, the lowest-order scattering amplitude, representing, for example, the
interaction of a spin-1/2 electron with a spin-0 monopole is
\be
T^{ss'}=2ieg\frac{(2\pi)^{4}\delta\left(p-p^{\prime}
+k-k^{\prime}\right)\epsilon_{\mu\nu\lambda\sigma}n^{\nu}
{\bar v}^{s^{\prime}}\left(p^{\prime}\right)
\gamma^{\mu}u^{s}(p)q^{\lambda}
\left(k+k^{\prime}\right)^{\sigma}}
{\sqrt{2E_{p}2E_{p^{\prime}}2\omega_{k}2\omega_{k^{\prime}}} 
\left(q^2 - i\epsilon\right)
\left(n\cdot q + i\delta\right)}\Bigg|_{q=p-p^{\prime}}.
\ee
Here the incoming
momenta are $p$, $k$, and the outgoing momenta are $p'$, $k'$, respectively,
while the initial and final electron spin projections are $s$ and $s'$, 
and where the different symbols for the energies of the electron and monopole,
$E$ and $\omega$, respectively,
refer to the different masses of the corresponding particles.
Further, note that
in this paper we use a metric with signature $(-,+,+,+)$, so that $q^2>0$
represents a spacelike momentum transfer.
The explicit dependence on the covariant 
string vector, $n_\mu$, {\em does not disappear\/} upon 
squaring the amplitude and multiplying by the appropriate phase space factors.
To make matters worse, the value of the magnetic charge
implied by (\ref{quan}),  $\alpha_{g}=g^2/{4\pi}\approx 34N^2$, calls 
into question any approach based on a badly divergent
perturbative expansion in $\alpha_{g}$.  Although these earliest efforts using 
a Feynman-rule perturbation theory in one of a number of
approaches to dual QED  (see below) resulted in string-dependent 
cross--sections \cite{wei65,rab69}, subsequently {\em ad hoc\/} 
assumptions were
invoked to render the resulting cross--sections string independent 
(\eg Refs.~\cite{dea82} and \cite{car96}).  

Recently, in the context of calculating virtual monopole processes,
De R\'{u}jula\cite{der95} has advocated the notion
that single photon-mediated monopole-antimonopole Drell-Yan production 
amplitudes are rendered string independent through the implausible 
argument of  Deans\cite{dea82}, which amounts to dropping the pole terms in
the photon propagator!\footnote{
Essentially it amounts to discarding string-dependent contributions
on the basis of the argument that
they cannot contribute to any gauge invariant quantity if the
theory is to be viable.}  Further, in a series of  papers,
Ignatiev and Joshi\cite{ign97}, 
while arguing that the prescription of Deans lacks believability,
propose (as did Rabl \cite{rab69})  averaging 
the scattering amplitude over all possible directions of the string  in order 
to eliminate the string dependence.  In contrast to Deans, however,
they arrive at a null result for the amplitude in lowest order.  
Consequently, they argue that general principles of quantum field theory 
which do not rely upon the use of perturbation theory must be invoked; that is,
they consider the constraints placed upon the scattering amplitude 
due to the requirements of discrete C and P 
symmetries (see also Tolkachev et al.~\cite{tom}).  

In contrast, by studying the {\it formal\/} behavior of Green's functions
in the relativistic quantum field theory of electrons and monopoles, both 
Schwinger \cite{sch68,sch75} and Brandt, Neri and 
Zwanziger\cite{zwa78} demonstrated Lorentz and gauge invariance.  
In essence these demonstrations are 
nonperturbative and  rely on the use of pointlike  particle
trajectories for the matter fields.  In the Schwinger
approach, however, classical particle  currents 
\be
j^\mu(x)&=&\sum_{e}e\int\, dx_{e}^\mu\delta\left(x-x_e\right)
\nn
\jg^\mu(x)&=&\sum_{g}g\int\, dx_{g}^\mu\delta\left(x-x_g\right)\
\label{ptsrc}
\ee
are  substituted for the field theoretic sources,
whereupon a change in the string trajectory
gives rise to a change in the action which is a multiple
of $2\pi$ because of the quantization conditions (\ref{quan}) or (\ref{dyon}).
In the second approach, putting aside the issue of renormalization, 
Brandt {\em et al.}\ succeed in  demonstrating the string 
independence of correlation functions of gauge- and Lorentz-invariant 
quantities using a functional path integral formalism,
once again as a consequence of Eqs.~(\ref{quan}) and (\ref{dyon}),
by  converting the path integral 
over fields to one over closed particle trajectories 
(see Refs.~\cite{fey50,sch51,hal77}).

However, while Lorentz  invariance and string independence
were {\it formally\/} demonstrated in dual QED (albeit 
with the above caveats), such demonstrations have been conspicuously 
absent in  practical calculations.\footnote{This is  surprising 
because one expects that the invariant  nonrelativistic scattering result 
(see Ref.~\cite{mil76} and references therein)
corresponds in a certain kinematic regime to a infinite summation of a 
particular {\em subclass} of Feynman diagrams.
Moreover, the relativistic Dirac equation also yields a string-independent
cross section \cite{kazama}.}
This deficiency  stems from the fact that in most 
phenomenological treatments of  
electron-monopole processes the ``string independence'' of 
the quantum field theory and the strength of the coupling are treated as 
separate issues.  This viewpoint (see \eg Refs.~\cite{ign97,car96})
could not be more misleading. 
In fact, these two points are intimately 
related.\footnote{In this regard we remind the reader of the 
following viewpoint of Schwinger: ``Relativistic invariance 
will appear to be violated in any treatment based on a perturbation 
expansion.  Field theory is more than a set 
of `Feynman's rules'.''\cite{sch66}}  If any lesson is to be learned 
from the above-mentioned demonstrations of Lorentz and 
string invariance, it is this: The quantization conditions (\ref{quan}) 
and (\ref{dyon}),  being  intimately tied to the demonstration of 
Lorentz invariance, are intrinsically nonperturbative statements (and 
for that matter are obtained independently of any perturbative 
approximation).  As such, attempts to demonstrate 
Lorentz invariance, whether formally or phenomenologically, which are based 
on a perturbative expansion in the coupling will most assuredly fail.
While this is not a novel  viewpoint (see Refs.~\cite{sch68,zwa78}),
it seems to have been overlooked in the 
more recent studies of point monopole  processes 
(see also Ref.~\cite{car96}).\footnote{A very recent note \cite{LN} exemplifying
the failure to connect the issue of the strength of the magnetic coupling
with that of string independence deduces the inverse renormalization of
electric and magnetic charges in contradistinction to the identity of
these renormalizations shown by Schwinger \cite{sch66r}.}

In fact, there exists one instance in the
literature of a successful calculation demonstrating string 
independence for a relativistic scattering cross section 
within dual QED.
Utilizing Schwinger's functional source theory \cite{sch66b,sch68} 
in the context of a high energy, low momentum transfer 
(zeroth order eikonal) approximation ({\it e.g.} 
Refs.~\cite{fri65,lev69,ab69,dit70,che87} and references therein), 
Urrutia \cite{urr78}  demonstrated string independence of the 
charge-monopole scattering cross section, although again
in his treatment the currents were approximated by those of
classical point particles 
as in Eq.~(\ref{ptsrc}).

The outline of this paper is as follows.
In Section II we establish the string-dependent dual-electrodynamic action.
In Section III we quantize the dual potential action by
solving the Schwinger-Dyson equations for the vacuum
persistence amplitude for electron-monopole (and dyon-dyon)
processes.  Then, by taking the functional Fourier transform
of this solution, we are led unambiguously to the dual QED 
functional path integral  that is compatible
with the source theory formalism developed by
Schwinger \cite{sch68,sch75}. In the process of this
development, we enforce a gauge fixing
condition that suggests that using the Dirac string formulation
is not only entirely natural and consistent with 
the underlying gauge symmetry of the charge-monopole
system, but in fact is preferable to the
multi-valued potential of Ref.~\cite{wu75}.
Having accomplished this,  in Section IV
we calculate the dyon-dyon scattering cross section
in the context of this relativistic {\it string dependent\/}
version of dual QED and demonstrate that a nonperturbative 
approach in conjunction with (\ref{quan}) 
or (\ref{dyon}) is in fact  necessary  to
demonstrate Lorentz and gauge (string) invariance of 
physical observables.
To accomplish this we use the nonperturbative functional field-theory 
formalism of  Schwinger \cite{sch51} and Symanzik \cite{sym54}.\footnote{For 
and excellent review of these techniques see \cite{fri90}.}
Within this formulation we are able to generalize the string-independent
eikonal result of Urrutia \cite{urr78} while treating the currents as
constructed from quantum fields 
by invoking the quantization condition (\ref{dyon}).
Finally in Section V we draw some conclusions concerning the 
phenomenology of 
dual QED and outline our continuing efforts in this direction while in 
addition proposing future work. Appendices deal with some aspects of the
path-integral formalism in dual QED.

\section{Dual  Electrodynamics}

Decades after Dirac's early quantum-mechanical  treatment 
(both nonrelativistic \cite{dir31} and relativistic\cite{dir48}) 
of the electron-monopole system, attempts to establish a
consistent interacting quantum field theory of (point) electrons
and monopoles were carried out in two different approaches by Schwinger
\cite{sch66} and by Zwanziger \cite{zwa71}.

Schwinger's approach is based upon the construction of a {\em nonlocal} 
Hamiltonian which is a function of  two transverse, string-dependent
potentials, $\bbox{A}^{\rm T}$ and $\bbox{B}^{\rm T}$,
in addition to, say, spin-$1/2$ electron and monopole fields.  
Second quantization is 
established by postulating the commutation rules between the fields and their 
conjugate field momenta.  In this formulation the two gauge 
fields have a nonvanishing commutator, and in fact are not 
{\em linearly independent}.  Although this operator 
method is noncovariant, it has the merit of explicitly 
displaying the dual symmetry between the Dirac equations for the 
spin-$\frac{1}{2}$ electron and monopole fields, as 
well as of the Maxwell equations for the magnetic and electric fields.
In turn Schwinger \cite{sch68,sch75}  developed and advocated 
a covariant formalism derived from his source theory
approach to quantum field theory.  This {\it first-order\/} formulation of 
dual QED is written in terms of the field strength tensor $F_{\mu\nu}$
and one independent vector potential 
$A_{\mu}$.  Alternatively, one can trade the dependence on the field
strength for that on an auxiliary dual vector potential
$B_\mu$.  The latter field is introduced to 
identify unambiguously the photon field
coupling to the dual monopole current, 
$\jg$, through the effective ``skeletal'' action (\ref{interact}),
coupling electric and magnetic currents.
This latter formulation proves helpful in calculating scattering processes. 

A second {\em local} covariant Lagrangian formulation approach, 
developed by Zwanziger, utilizes two 
four-potentials $A_{\mu}$ and $B_{\mu}$, which
again are not independent. In addition, this formalism
satisfies  duality symmetry both at the level of the action as well as
between the equations of motion for $ A_{\mu}$ and $B_{\mu}$ 
and between the equations 
for the spin-$\frac{1}{2}$ electron and monopole fields.
The equivalence between these formalisms is shown in Refs.~\cite{BN,bla78}.

However, the former covariant  formalism of Schwinger is 
a natural generalization of Dirac's (relativistic) 
quantum mechanics in the form of a covariant  Lagrangian field 
theoretic description of electron-monopole interactions.
In this context it should come as no surprise that Schwinger's source theory
approach \cite{sch75} laid the groundwork for the first 
consistent, gauge-invariant nonperturbative calculation of 
electron-monopole scattering \cite{urr78}.  

We also note that a ``one potential'' Hamiltonian
formulation was developed later 
by Blagojevi\'{c} \cite{bla78} where a {\it explicit\/} 
set of Feynman rules was displayed.   Finally we mention in passing the 
work of Deans \cite{dea82}, where a formalism based on Dirac's monopole
theory was quantized within the framework of Mandelstam's gauge-invariant
quantum field theory~\cite{man62}.  In Deans' paper, while  a set 
of {\it string-dependent\/} Feynman rules was developed, they were used
without criticism of the badly divergent 
perturbation series; in addition, as we have seen, Lorentz-frame (string)
dependence is argued away in a highly questionable fashion.

\subsection{ The ``Dual-Potential'' String  Dependent Action}
\label{ss:onepot}

While the completely general formalism for charge-monopole
quantum field theory was developed by Schwinger in the context 
of source theory, for the sake of accessibility here we develop a
string-dependent dual QED of monopole-electron interactions
in a more familiar functional  formalism.\footnote{However, 
we emphasize that the source theory
approach is in fact just that, the ``source'' of these ideas.}

In order to facilitate the construction of the dual-QED 
formalism we recognize that the well-known 
continuous global ${\rm U}(1)$ {\it dual} symmetry \cite{sch66,sch68,sch75}
implied by Eqs.~(\ref{max}), (\ref{con}), given by 
\begin{mathletters}
\be
\left( \begin{array}{c} 
j^{\prime}  \\  {\jg}^{\prime}
\end{array} \right)
&=&\left( \begin{array}{cc} \cos\phi &  \sin\phi \\  
-\sin\phi & \cos\phi  
\end{array} \right) 
\left( \begin{array}{c} j  \\  {\jg} \end{array} \right),
\\
\left( \begin{array}{c} 
F^{\prime} \\ {\fg}^{\prime} 
\end{array} \right) 
&=&\left( \begin{array}{cc} \cos\phi & \sin\phi  \\  
-\sin\phi & \cos\phi  
\end{array} \right)
\left( \begin{array}{c} F  \\  {\fg} \end{array} \right)\, ,
\ee
\end{mathletters}
suggests the introduction of an auxiliary vector potential $B_\mu(x)$ 
dual to $A_\mu(x)$. 
In order to satisfy the Maxwell and charge conservation  
equations, Dirac  modified the field strength tensor according to
\be
F_{\mu\nu}=\partial_{\mu}A_{\nu}-\partial_{\nu}A_{\mu}+\Gg_{\mu\nu}\, ,
\label{fs}
\ee
where now Eqs.~(\ref{max}) and (\ref{con}) give rise to the
consistency condition on $G_{\mu\nu}(x)=-G_{\nu\mu}(x)$
\be
\partial^{\nu}\ ^{\ast}F_{\mu\nu}=-\partial^{\nu}
G_{\mu\nu}={\jg}_\mu\, .  
\label{subc}
\ee
We then obtain the following inhomogeneous solution
to the dual Maxwell's equation (\ref{subc}) for the tensor
$G_{\mu\nu}(x)$ in terms of the string function $f_\mu$ and the magnetic 
current, which for a spin-1/2 monopole represented by a Dirac field $\chi$
is $\jg^{\mu}(x)=g\bar{\chi}(x)\gamma^\mu\chi(x)$:
\be
G_{\mu\nu}(x)&=&\left(n\cdot\partial\right)^{-1}
\left(n^{\mu}\jg^{\nu}(x)-n^{\nu}\jg^{\mu}(x)\right)
\nn
&=&\int \dy \left(f_{\mu}(x-y)\jg_{\nu}(y)
-f_{\nu}(x-y)\jg_{\mu}(y)\right)\, ,
\label{gten}
\ee
where use is made of Eqs.~(\ref{con}), (\ref{dfstng}), and (\ref{stngsl}).
A minimal generalization of the QED Lagrangian including
electron-monopole interactions reads
\be
\cal L&=&-\frac{1}{4}F_{\mu\nu}F^{\mu\nu}
+\bar\psi \left(i\dslash+e\Aslash-m_{\psi}\right)\psi
+\bar\chi \left(i\dslash-m_{\chi}\right)\chi
\label{lag0}\, ,
\ee
where the coupling of the monopole field $\chi(x)$ to the electromagnetic
field occurs through the quadratic field strength term according to
Eq.~(\ref{fs}).  
We now rewrite the Lagrangian~(\ref{lag0})
to  display more clearly that interaction
by introducing the auxiliary potential $B_\mu(x)$.

Variation of Eq.~(\ref{lag0}) with respect to the field variables,
$\psi$, $\chi$ and $A_\mu$, yields in addition to the Maxwell 
equations for the field strength,\footnote{We regard $G_{\mu\nu}(x)$
as dependent on $\bar{\chi}$, $\chi$ but not $A_\mu$.  Thus,
the dual Maxwell equation is given by the subsidiary condition~(\ref{subc}).} 
$F_{\mu\nu}$, Eq.~(\ref{max}) where 
$j^{\mu}(x)=e\bar{\psi}(x)\gamma^\mu\psi(x)$,
the equation of motion for the electron field
\be
\left(i\dslash + e\Aslash(x) - m_{e}\right)\psi(x) =0\, ,
\label{diracl}
\ee
and the nonlocal equation of motion for the monopole field,
\be
\left(i\dslash - m_{\chi}\right)\chi(x)-\frac{1}{2}\int \left(dy\right) 
\fg^{\mu\nu}(y)\frac{\delta G_{\mu\nu}(y)}{\delta\bar\chi(x)} = 0\, .
\label{diracm}
\ee
It is straightforward to see from the Dirac 
equation for the monopole (\ref{diracm}) and the construction (\ref{gten})
that introducing the auxiliary dual field (which is a 
functional  of $F_{\mu\nu}$ and depends on the string function $f_\mu$)
\be
B_{\mu}(x)=-\int \dy f^\nu\left(x-y\right)\fg_{\mu\nu}(y)\, ,
\label{gfb}
\ee
results in the following Dirac equation for the monopole field
\be
\left(i\dslash +g\Bslash(x) - m_{g}\right)\chi(x)&=&0\, .
\label{diracm2}
\ee
Here we have chosen the string to satisfy the oddness condition
(this is the ``symmetric'' solution)
\be
f^\mu(x)=-f^\mu(-x),
\ee
which is related to Schwinger's integer quantization condition \cite{mil76,psf}.
Now (\ref{diracl}) and (\ref{diracm2})
display the dual symmetry expressed in Maxwell's equations
(\ref{max}) and (\ref{con}). 
Noting that  $B_\mu$   satisfies
\be
\int (d x') f^{\mu}(x-x^{\prime})B_{\mu}(x^{\prime})=0\, ,
\ee
we see that Eq.~(\ref{gfb}) is a gauge-fixed vector field \cite{dir55,sis87} 
defined in terms of the field strength through an {\em inversion}
formula (see Subsection \ref{ss:gauge}). 
In terms of these fields the ``dual-potential''   action 
can be re-expressed in terms of the vector potential $A_\mu$ 
and field strength tensor $F_{\mu\nu}$ (where $B_\mu$ is the functional
(\ref{gfb}) of $F_{\mu\nu}$) in first-order formalism as 
\begin{mathletters}
\be
W&=&\int\dx \bigg\{-\frac{1}{2}F^{\mu\nu}(x)
\left(\partial_\mu A_\nu\left(x\right)-\partial_\nu A_\mu\left(x\right)\right)+
\frac{1}{4}F_{\mu\nu}(x)F^{\mu\nu}(x)
\nn
&+& 
\bar\psi(x)\left(i\dslash+e{\Aslash}(x) 
-m_{\psi}\right)\psi(x)
+\bar\chi(x)
\left(i\dslash+g{\Bslash}(x)-m_{\chi}\right)\chi(x)\bigg\}\, ,
\label{act1}
\ee
or in terms of {\em dual\/} variables,
\be
W&=&\int \dx \bigg\{
-\frac{1}{2}{\fg}^{\mu\nu}(x)\left(\partial_\mu B_\nu\left(x\right)
-\partial_\nu B_\mu\left(x\right)\right)
+\frac{1}{4}{\fg}^{\mu\nu}(x){\fg}_{\mu\nu}(x)
\nonumber \\
&+&
\bar\psi(x)\left(i\dslash+e{\Aslash}(x)-m_{\psi}\right)\psi(x)
+\bar\chi(x)\left(i\dslash+g{\Bslash}(x)-m_{\chi}\right)\chi(x) \bigg\}\, .
\label{act2}
\ee
\end{mathletters}
In Eq.~(\ref{act1}), $A_\mu(x)$ and $F_{\mu\nu}(x)$ are the independent 
field variables\footnote{Using Eq.~(\ref{gfb}), variations 
of the  action, Eq.~(\ref{act1}), with respect
to $A_\mu(x)$ and $F_{\mu\nu}(x)$ 
yield Eqs.~(\ref{max}) and (\ref{fs}) where $\Gg_{\mu\nu}(x)$ is 
the dual of Eq.~(\ref{gten}).} and $B_\mu(x)$ is given 
by Eq.~(\ref{gfb}),  while in Eq. (\ref{act2}) the dual fields 
are the independent variables, in which case,
\be
A_\mu(x)&=&-\int \dy f^\nu\left(x-y\right)\ F_{\mu\nu}(y)\nonumber\\
&=&{1\over2}\epsilon_{\mu\nu\lambda\sigma}\int (dy)f^\nu(x-y)\fg^{\lambda
\sigma}(y).
\label{gfa0}
\ee
[Note that Eq.~(\ref{act2}) may be obtained from the form (\ref{act1}) by
inserting Eq.~(\ref{gfa0}) into the former and then identifying $B_\mu$
according to the construction (\ref{gfb}).  In this way the sign of
${1\over4}F_{\mu\nu}F^{\mu\nu}=-{1\over4}{}^*F_{\mu\nu}{}^*F^{\mu\nu}$ is
flipped.]
Consequently, the field equation relating $\fg^{\mu\nu}$ and $B^\mu$ is
\be
{\fg}_{\mu\nu}=\partial_{\mu}{B}_{\nu}-\partial_{\nu}{B}_{\mu}
-\int \dy \,{}^{\ast}\left(f_{\mu}(x-y)j_{\nu}(y)
-f_{\nu}(x-y)j_{\mu}(y)\right),
\label{dualfs}
\ee
which is simply obtained from Eq.~(\ref{fs}) by making the duality
transformation ${\cal E}\to{\cal M}$, ${\cal M}\to-{\cal E}$, where
${\cal E}$ stands for any electric quantity and ${\cal M}$ for any
magnetic quantity.

\section{Quantization of Dual QED: Schwinger-Dyson Equations}
\label{ss:quant}

Although the various actions describing the interactions of
point electric and magnetic poles can be described in terms 
of a set of Feynman rules which one conventionally uses
in perturbative calculations, the large value of
$\alpha_{g}$ or $eg/4\pi$ renders them useless
for this purpose. In addition, as mentioned
in Section I, calculations of physical processes using
the perturbative approach from string-dependent actions 
such as Eqs.~(\ref{act1}) and (\ref{act2})
have led only to string dependent results.
In conjunction with a nonperturbative 
functional approach, however, the Feynman rules serve to elucidate 
the electron-monopole interactions.  We express these interactions
in terms of the ``dual-potential'' formalism 
as a quantum generalization of the relativistic 
classical theory of section \ref{ss:onepot}.
We use the Schwinger action principle \cite{sch51p} 
to quantize the electron-monopole  system by 
solving the corresponding Schwinger-Dyson equations for the 
generating functional. Using a functional Fourier
transform of this generating functional in terms 
of a path integral for the electron-monopole system,
we  rearrange the generating 
functional into a form that is well-suited for the purpose of 
nonperturbative calculations.

\subsection{Gauge Symmetry}
\label{ss:gauge}

In order to construct the generating functional for Green's
functions in the electron-monopole system we must restrict
the gauge freedom resulting from the local gauge invariance
of the action (\ref{act1}).
The {\it inversion} 
formulae for $A_\mu$ and  $B_\mu$, Eqs.~(\ref{gfa0}) and (\ref{gfb})
respectively, might suggest using the technique of gauge-fixed 
fields \cite{dir55,man62} as was adopted in \cite{dea82}. 
However, we use the technique of gauge 
fixing according to methods outlined by Zumino~\cite{zum60}
and generalized by Zinn-Justin~\cite{zin86} in the language of
stochastic quantization.

The gauge fields are obtained in terms of the string and the gauge 
invariant field strength, by contracting the field 
strength (\ref{fs}), (\ref{gten}) with the Dirac string, $f^{\mu}(x)$,
in conjunction with Eq.~(\ref{dfstng}), yielding 
the following inversion formula for the equation of motion, 
\be
A_\mu(x)&=&-\int \dxp f^{\nu}(x-x^{\prime})F_{\mu\nu}(x^{\prime})
+\partial_{\mu}\tilde{\Lambda}_{e}(x)
\label{sta}\ ,
\ee
where we use the suggestive notation, ${\tilde\Lambda}_{e}(x)$
\be 
\tilde{\Lambda}_{e}(x)
=\int \dxp f^{\nu}(x-x^{\prime}) A_\nu\left(x^{\prime}\right)\, .
\label{agf}
\ee
It is evident that Eq.~(\ref{sta}) transforms consistently under
 gauge transformation
\be
A_\nu(x)\longrightarrow A_\nu(x)+\partial_{\nu}\Lambda_{e}(x)\, ,
\ee
while in addition we note that the Lagrangian  (\ref{act1}) 
is invariant under the gauge transformation,
\begin{mathletters}
\be
&\psi & \rightarrow  {\exp}\left[ie\Lambda_{e}\right]\psi\, ,
\hspace{.25cm}
A_\mu \rightarrow  A_\mu + \partial_\mu\Lambda_{e}\, ,
\ee
as is the dual action (\ref{act2}) under
\be
&\chi &  \rightarrow  {\rm exp}\left[ig\Lambda_{g}\right]\chi\, ,
\hspace{.25cm}
B_\mu \rightarrow  B_\mu + \partial_\mu\Lambda_{g}\, .
\ee\end{mathletters}
Assuming the freedom to  choose $\tilde{\Lambda}_e(x)=-\Lambda_{e}(x)$,
we bring the vector potential into gauge-fixed form,\footnote{It is worth 
noting the similarity of this condition to the Schwinger-Fock gauge 
in ordinary QED, $x\cdot\A(x)=0$ which yields the gauge-fixed 
photon field, $\A_{\mu}(x)=-x^\nu\int_0^1 ds\,s F_{\mu\nu}(xs)$.}
coinciding with Eq.~(\ref{gfa0}),
\be
A_{\mu}(x)&=&-\int \dy f^\nu\left(x-y\right)\ F_{\mu\nu}(y)
\label{gfa}
\ee
where the gauge choice is equivalent 
to a {\it string-gauge\/} condition\footnote{
Taking the divergence of Eq.~(\ref{gfa}) and using Eq.~(\ref{max}),
the gauge-fixed condition (\ref{gfa}) can be written as
\be
\partial_\mu\A^\mu = \int \dy f^\mu\left(x-y\right) j_{\mu}(y)\, ,
\label{stgc1}
\ee
which is nothing other than the gauge-fixed condition of Zwanziger
in the two-potential formalism \cite{zwa71}.}  
\be
\int\ \dxp f^{\mu}(x-x^{\prime})A_{\mu}(x^{\prime})=0\, .
\label{gfca}
\ee
More generally, the fact that a gauge function exists,
such that 
${\Lambda}_e(x)=-\tilde\Lambda_e(x)$ [\cf\/ Eq. (\ref{agf})], implying 
that we have the freedom to consistently fix the gauge,
is in fact not a trivial claim. If this were not true, it 
would certainly derail the consistency
of incorporating monopoles into QED while 
utilizing the Dirac string  formalism.  On the contrary, the
{\it string gauge condition}, Eq.~(\ref{gfca}), is in fact 
a class of possible consistent gauge 
conditions characterized by the symbolic operator function (\ref{stngsl})
depending on a unit vector $n^\mu$ (which may be either spacelike or timelike).
In a similar manner, given the dual field strength (\ref{dualfs})
the dual vector potential takes the following form [{\cf} Eq. (\ref{gfb})]
\be
B_{\mu}(x)
=-\int\dxp f^{\nu}(x-x^{\prime}){\fg}_{\mu\nu}(x^{\prime})+ 
\partial_{\mu}\tilde{\Lambda}_{g}\, ,
\label{gfb2}
\ee
where
\be
\tilde{\Lambda}_{g}(x)=\int\dxp f^{\mu}(x-\xp)B_{\mu}(x')\, .
\label{bgf}
\ee

In order to quantize this system we must 
divide out the equivalence class of field values
defined by a gauge trajectory in field space; in this
sense the gauge condition restricts the
vector potential to a hypersurface of field space
which is embodied in the generalization of Eq.~(\ref{gfca})
\be
\int\dxp f^\mu(x-\xp)A_\mu(\xp)=\Lambda_{e}(x)\, ,
\label{stgc2}
\ee
where here $\Lambda_e$ is any function defining a unique gauge
fixing hypersurface in field space.\footnote{Choosing a different
function $\Lambda_e$ (which the gauge freedom permits us to do)
merely yields a different section of field space under the
restriction that it cut each equivalence class of field
values once.}

In a path integral formalism, we enforce the condition (\ref{stgc2})
by introducing a $\delta$ function, symbolically written as
\be
\delta(f^{\mu}A_{\mu}-\Lambda_e)&=&\int 
[d\lambda_e]\exp\left[i\int\dx\lambda_e(x)
\left(\int\dxp f^{\mu}(x-\xp)A_{\mu}(\xp)-\Lambda_e(x)\right)\right]\, ,
\label{delta}
\ee
or by introducing a Gaussian functional integral
\be
\Phi(f^{\mu}A_{\mu}-\Lambda_e)&=&\int 
[d\lambda_e]\exp\Bigg[-\frac{i}{2}\int\dx\dxp
\lambda_e(x)M(x,\xp)\lambda_e(x')
\nn
&&+i\int\dx\lambda_e(x)\left(\int\dxp f^{\mu}(x-\xp)A_{\mu}(\xp) 
-\Lambda_e(x)\right)\Bigg]\, ,
\label{gfcq}
\ee
where the symmetric matrix $M(x,\xp)=\kappa^{-1}\delta(x-\xp)$ 
describes the spread of the
integral $\int\dxp f^{\mu}(x-\xp)A_{\mu}(\xp)$ about the gauge
function, $\Lambda_e(x)$. That is, we  enforce the gauge fixing
condition (\ref{stgc2}) by adding the quadratic form appearing here
to the action (\ref{act1}) and in turn eliminating
$\lambda_{e}$ by its ``equation of motion''
\be
\lambda_{e}(x)=\kappa\left(\int (dy)\,f^\mu(x-y)
A_\mu(y)-\Lambda_e(x)\right)\, .
\label{agfl}
\ee
Now the equations of motion (\ref{max}) take the form,
\begin{mathletters}
\be
\partial^\nu F_{\mu\nu}(x)&-&\int(dx') \lambda_e(\xp)f_\mu(\xp-x)=j_\mu(x),
\label{gfmax1}
\\ 
\partial^\nu \fg_{\mu\nu}(x)&-&\int(dx') \lambda_g(\xp)f_\mu(\xp-x)=\jg_\mu(x).
\label{gfmax2}
\ee
\end{mathletters}
where the second equation refers to a similar gauge fixing in the dual sector.
Taking the divergence of Eqs.~(\ref{gfmax1}) implies $\lambda_e=0$
from Eqs.~(\ref{dfstng}) and (\ref{con}), which
consistently yields the gauge condition (\ref{stgc2}).
Using our freedom to make a transformation to the gauge-fixed condition 
(\ref{gfa}), $\Lambda_e=0$, the equation of motion (\ref{gfmax1}) 
for the potential becomes
\be
\Bigg[-g_{\mu\nu}\partial^2 + 
\partial_\mu\partial_\nu +\kappa ~n_{\mu}(n\cdot\partial)^{-2}n_{\nu}
\Bigg]A^\nu(x)
= j_\mu(x) +\epsilon_{\mu\nu\sigma\tau}
\frac{n^\nu}{\left(n\cdot\partial\right)}\partial^{\sigma}
\jg^\tau(x)\, ,
\quad n^\mu A_\mu=0,\nonumber\\
\label{difa2}
\ee
where we now have used the symbolic form of the string function~(\ref{stngsl}).
We have retained the term proportional to $n_\mu n_\nu$ in
 the  kernel, scaled by the arbitrary parameter $\kappa$,
\be
K_{\mu\nu}=
\Bigg[-g_{\mu\nu}\partial^2 + 
\partial_\mu\partial_\nu+\kappa\, n_{\mu}(n\cdot\partial)^{-2}n_{\nu}
\Bigg]
\label{ker}
\ee
so that $K_{\mu\nu}$ possesses an inverse 
\be
D_{\mu\nu}(x)&=&\Bigg[g_{\mu\nu}
-\frac{n_{\mu}\partial_{\nu}+n_{\nu}\partial_{\mu}}{(n\cdot\partial)}
+n^{2}\left(1-\frac{1}{\kappa}\frac{(n\cdot\partial)^{2}\partial^{2}}{n^{2}}\right)
\frac{\partial_{\mu}\partial_{\nu}}{(n\cdot\partial)^{2}}\Bigg]D_+(x)\, ,
\ee
that is, $\int\dxp K_{\mu\alpha}(x-\xp)D^{\alpha\nu}(\xp-\xpp)=
g_\mu^\nu\delta(x-x'')$,
where $D_+(x)$ is the massless scalar propagator,
\be
D_+(x)=\frac{1}{-\partial^{2}-i\epsilon}\delta(x)\, .
\ee
This in turn enables us to rewrite Eq. (\ref{difa2}) as 
an integral equation, expressing the vector potential in terms of the
electron and monopole currents,
\be
A_{\mu}(x)&=&\int\dxp D_{\mu\nu}(x-\xp)j^{\nu}(\xp)
\nn
&& \mbox{}+\epsilon^{\nu\lambda\sigma\tau}\int\dxp\dxpp 
D_{\mu\nu}(x-\xp)f_{\lambda}(\xp-\xpp)\partial''_{\sigma}\jg_{\tau}(\xpp)\, .
\label{inta}
\ee
The steps for $B_\mu(x)$ are analogous.

\subsection{Vacuum Persistence Amplitude and the Path Integral}
\label{ss:vpa}

Given the gauge-fixed but string-dependent action we
are prepared to quantize this theory of dual QED.
Quantization using a path integral formulation of such
a string dependent action is by no means straightforward;
therefore we will develop the generating functional making use of
a functional approach.
Using the quantum action principle
(\cf\ Ref.~\cite{sch51p}) we write the generating functional 
for Green functions (or the vacuum persistence 
amplitude)   in the presence of external sources~$\J$
\be
Z({\cal J})&=&
\langle 0_+\left|\right. 0_-\rangle^{\cal J}\, ,
\ee
for the electron-monopole system.
Schwinger's action principle
states that under an arbitrary variation
\be
\delta\langle 0_+\left|\right. 0_-\rangle^{\J}&=&
i\langle 0_+\left| \delta W(\J)\right| 0_-\rangle^{\J}\, ,
\ee
where $W(\J)$ is the action given in Eq.~(\ref{act1}) externally 
driven by the sources, $\J$, which for the present case are
given by the set $\{J,\Jg,\bar{\eta},\eta,\bar{\xi},\xi\}$:
\be
W({\cal J})=W+\int (dx)\left\{J^\mu A_\mu+{}^* J^\mu B_\mu+\bar\eta\psi
+\bar\psi\eta+\bar\xi\chi+\bar\chi\xi\right\}.
\ee
The one-point functions are then given by
\be
\frac{\delta}{i\delta J^\mu(x)}
\log Z(\J)&=&
\frac{\langle 0_+|A_\mu(x)| 0_-\rangle^\J}
{\langle 0_+\left|\right. 0_-\rangle^{\cal J}}\, , \hspace{1cm} 
\frac{\delta}{i\delta\Jg^\mu(x)}
\log Z(\J)=
\frac{\langle 0_+|B_\mu(x)| 0_-\rangle^\J}
{\langle 0_+\left|\right. 0_-\rangle^{\cal J}}\, ,
\nn
\frac{\delta}{i\delta\bar{\eta}(x)}
\log Z(\J)&=&
\frac{\langle 0_+| \psi(x)| 0_-\rangle^\J}
{\langle 0_+\left|\right. 0_-\rangle^{\cal J}}\, , \hspace{1cm} 
\frac{\delta}{i\delta\bar{\xi}(x)}
\log Z(\J)=
\frac{\langle 0_+|\chi(x)| 0_-\rangle^\J}
{\langle 0_+\left|\right. 0_-\rangle^{\cal J}}\, .
\label{onept}
\ee
Using Eqs.~(\ref{onept}) we can write down
derivatives with respect to the charges\footnote{Here 
we redefine the electric and magnetic currents
$j\rightarrow ej$ and $\jg\rightarrow g\jg$. Note that the changes
in the action due to induced changes in the fields vanish by virtue of
the stationary principle.} 
in terms of functional derivatives~\cite{sch61,som63,jon65} 
with respect to the external sources;
\be
\frac{\partial}{\partial e}
\langle 0_+| 0_-\rangle^{\J}
&=&i\langle 0_+\big|\int\dx j^\mu(x)A_\mu(x)\big| 0_-\rangle^{\J}
\nn
&=&-i\int\dx\left(\frac{\delta}{\delta\tA_\mu(x)}
\frac{\delta}{\delta J^\mu(x)}\right)
\langle 0_+\big| 0_-\rangle^{\J},
\nn
\frac{\partial}{\partial g}\langle 0_+| 0_-\rangle^{\J}&=&
i\langle 0_+\big|\int\dx \jg^\mu(x)B_\mu(x)\big| 0_-\rangle^{\J}
\nn
&=&-i\int\dx\left(\frac{\delta}{\delta\tB_\mu(x)}
\frac{\delta}{\delta \Jg^\mu(x)}\right)
\langle 0_+\big| 0_-\rangle^{\J}\, .
\ee
Here we have introduced an effective source to bring down the electron
and monopole currents,
\be
{\delta\over\delta \tA_\mu}\equiv{1\over i}{\delta\over\delta \eta}\gamma^\mu
{\delta\over\delta \bar \eta},\quad
{\delta\over\delta \tB_\mu}\equiv{1\over i}{\delta\over\delta \xi}\gamma^\mu
{\delta\over\delta \bar \xi}.
\label{effcurrsource}
\ee
These first order differential equations can be integrated with the result
\be
\langle 0_+\left|\right. 0_-\rangle^{\J}=
\exp\left[-ig\int\dx\left(\frac{\delta}{\delta\tB_\nu(x)}
\frac{\delta}{\delta\Jg^\nu(x)}\right)
-ie\int\dx\left(\frac{\delta}{\delta\tA_\mu(x)}
\frac{\delta}{\delta J^\mu(x)}\right)\right]
\langle 0_+\left|\right. 0_-\rangle^{\J}_0\, ,
\label{vacint}
\ee
where $\langle 0_+\left|\right. 0_-\rangle^{\J}_0$ is the vacuum 
amplitude in the absence of interactions.
By construction, the vacuum amplitude and Green's functions 
for the coupled problem are determined by functional derivatives
with respect to the external sources $\J$ of the uncoupled vacuum amplitude,
where $\langle 0_+\left|\right. 0_-\rangle^{\J}_0$ is the product
of the separate amplitudes for the quantized electromagnetic
and Dirac fields since they constitute completely 
independent systems in the absence of coupling, that is,
\be
\langle 0_+\left|\right. 0_-\rangle^{\J}_0=
\langle 0_+\left|\right. 0_-\rangle^{(\bar\eta,\eta,\bar\xi,\xi)}_{0}
\langle 0_+\left|\right. 0_-\rangle^{(J,\Jg)}_{0}\, .
\label{fulfregen}
\ee
First we consider $\langle 0_+\left|\right. 0_-\rangle^{\J}_0$
as a function of $J$ and $\Jg$
\be
\frac{\delta}{i\delta J^\mu(x)}\langle 0_+\left|\right. 0_-\rangle^\J_{0}&=&
\langle 0_+| A_\mu(x)| 0_-\rangle^\J_{0}\, .
\ee
Taking the matrix element of the integral 
equation (\ref{inta}) but now with external sources rather than
dynamical currents we find
\be
\langle 0_+| A_\mu(x)| 0_-\rangle^\J_0&=&
\int\dxp D_{\mu\nu}(x-\xp)\left(J^{\nu}(\xp)
+\epsilon^{\nu\lambda\sigma\tau}\int\dxpp 
f_{\lambda}(\xp-\xpp)\partial''_{\sigma}
\Jg_{\tau}(\xpp)\right)\langle 0_+|0_-\rangle_0^{\cal J}\, .\nn
&&
\label{qinta}
\ee
Using Eq. (\ref{difa2}) we arrive at the equivalent
gauge-fixed functional equation,
\be
\Bigg[-g_{\mu\nu}\partial^2
+\partial_{\mu}\partial_{\nu}
&+&\kappa n_{\mu}(n\cdot\partial)^{-2}n_{\nu}
\Bigg]\frac{\delta}{i\delta J^{\nu}(x)}
\langle 0_+\left|\right. 0_-\rangle^{\J}_0
\nn
&&\hspace{3cm}=\left(J_{\mu}(x)+\epsilon_{\mu\nu\sigma\tau}\int\dxp 
f^{\nu}(x-\xp)\partial^{\prime\sigma}\Jg^{\tau}(\xp)\right)
\langle 0_+\left|\right. 0_-\rangle^{\J}_0\, ,
\label{qeqaf}
\ee
which is subject to the gauge condition
\begin{mathletters}
\be
n^\nu{\delta\over\delta J^\nu}\langle 0_+|0_-\rangle_0^{\cal J}=0,
\ee
or
\be
\int (dx')f^\nu(x-x'){\delta\over\delta J^\nu(x')}
\langle 0_+|0_-\rangle_0^{\cal J}=0.
\label{gcone}
\ee
\end{mathletters}
In turn, from Eq.~(\ref{vacint})  we obtain the full functional equation for
$\langle 0_+\left|\right. 0_-\rangle^{\J}$:
\be
\Bigg[-g_{\mu\nu}\partial^2
+\partial_{\mu}\partial_{\nu}
&+&\kappa~n_{\mu}(n\cdot\partial)^{-2}n_{\nu}
\Bigg]\frac{\delta}{i\delta J^\nu(x)}\langle 0_+\left|\right. 0_-\rangle^{\J}
\nn
&=&\exp\left[-ig\int\dy
\left(\frac{\delta}{\delta\tB_\alpha(y)}
\frac{\delta}{\delta \Jg^\alpha(y)}\right)
-ie\int\dy
\left(\frac{\delta}{\delta\tA_\alpha(y)}
\frac{\delta}{\delta J^\alpha(y)}\right)\right]
\nn
&&\times\left(
J_{\mu}(x) +\epsilon_{\mu\nu\sigma\tau}\int\dxp 
f^{\nu}(x-\xp)\partial^{\prime\sigma}\Jg^{\tau}(\xp)\right)
\langle 0_+\left|\right. 0_-\rangle^{\J}_0\, .
\label{fncteq1}
\ee
Commuting the external currents to the left of the exponential on
the right side of Eq.~(\ref{fncteq1})
and using  Eqs.~(\ref{onept}),
we are led to the Schwinger-Dyson equation for the vacuum amplitude,
where we have restored the meaning of the functional derivatives with
respect to $\tilde A$, $\tilde B$ given in Eq.~(\ref{effcurrsource}),
\be
\Bigg\{\bigg[&-&g_{\mu\nu}\partial^2
+\partial_{\mu}\partial_{\nu}
+\kappa n_\mu(n\cdot\partial)^{-2}n_\nu
\bigg]\frac{\delta}{i\delta J_\nu(x)}
\nn
&&- e\frac{\delta}{i\delta\eta(x)}\gamma_\mu\frac{\delta}{i\delta\bar\eta(x)}
-\epsilon_{\mu\nu\sigma\tau}\int \dxp
f^{\nu}(x-\xp)\partial^{\prime\sigma}g\frac{\delta}{i\delta\xi(x')}\gamma^{\tau}
\frac{\delta}{i\delta\bar\xi(x')}
\Bigg\}\langle 0_+\left|\right. 0_-\rangle^{\J}\nn
&&=\left(J_\mu(x)+\epsilon_{\mu\nu\sigma\tau}\int(dx')f^\nu(x-x')\partial^{
\prime\sigma}{}^*J^\tau(x')\right)\langle 0_+\left|\right. 0_-\rangle^{\J}
\label{dsi1}\, .
\ee
In an analogous manner, using
\be
\frac{\delta}{i\delta \Jg^\mu(x)}\langle 0_+\left|\right. 0_-\rangle^\J_{0}&=&
\langle 0_+| B_\mu(x)| 0_-\rangle^\J_{0}\, ,
\ee
we obtain the functional  equation (which is consistent with duality)
\be
\Bigg\{\bigg[&-&g_{\mu\nu}\partial^2
+\partial_{\mu}\partial_{\nu}
+\kappa n_{\mu}(n\cdot\partial)^{-2}n_{\nu}
\bigg]\frac{\delta}{i\delta \Jg_\nu(x)}
\nn
&&- g\frac{\delta}{i\delta\xi(x)}\gamma_\mu
\frac{\delta}{i\delta\bar\xi(x)}
+\epsilon_{\mu\nu\sigma\tau}\int \dxp
f^{\nu}(x-\xp)\partial^{\prime\sigma}
e\frac{\delta}{i\delta\eta(x')}\gamma^{\tau}
\frac{\delta}{i\delta\bar\eta(x')}
\Bigg\}\langle 0_+\left|\right. 0_-\rangle^{\J}
\nn
&&=\left(
\Jg_\mu(x)-\epsilon_{\mu\nu\sigma\tau}\int(dx')f^\nu(x-x')\partial^{\prime
\sigma} J^\tau(x')\right)\langle 0_+\left|\right. 0_-\rangle^{\J}
\label{dsi2}\, ,
\ee
which is subject to the gauge condition
\be
\int(dx')f^\mu(x-x'){\delta\over\delta {}^*J_\mu(x')}\langle 0_+|0_-\rangle^{
\cal J}=0.
\label{gctwo}
\ee
\begin{mathletters}
In a straightforward manner we obtain the functional Dirac equations
\be
\left\{i\dslash+e\gamma^\mu\frac{\delta}{i\delta J^\mu(x)}
-m_\psi\right\}
\frac{\delta}{i\delta\bar{\eta}(x)}\langle 0_+\left|\right. 0_-\rangle^{\J}
&=&-\eta(x)\langle 0_+\left|\right. 0_-\rangle^{\J}
\label{dsi3}\, ,\\
\left\{i\dslash+g\gamma^\mu\frac{\delta}{i\delta\Jg^\mu(x)}
-m_\chi\right\}
\frac{\delta}{i\delta\bar{\xi}(x)}
\langle 0_+\left|\right. 0_-\rangle^{\J}&=&-\xi(x)
\langle 0_+\left|\right. 0_-\rangle^{\J}
\label{dsi4}\, .
\ee
\end{mathletters}

In order to obtain a generating functional for Green's functions
we must solve the set of 
equations~(\ref{dsi1}), (\ref{dsi2}), (\ref{dsi3}), (\ref{dsi4}) 
subject to Eqs.~(\ref{gcone}) and (\ref{gctwo}) for 
$\langle 0_+\left|\right. 0_-\rangle^{\J}$.
In the absence of interactions, we can immediately
integrate the  Schwinger-Dyson equations;
in particular, (\ref{dsi2}) then integrates to
\be
\langle 0_+\left|\right. 0_-\rangle^{\J}_{0}
&=&\N\left(J\right)\exp\bigg\{
\frac{i}{2}\int(dx)(dx')\Jg_\mu\left(x\right)D^{\mu\nu}
(x-\xp)\Jg_\nu(\xp)
\nonumber \\
&&
\mbox{}+ i\epsilon_{\mu\nu\sigma\tau}
\int (dx)(dx')(dx'')\Jg_\beta(x)D^{\beta\mu}(x-\xp)\partial^{\prime\nu}
f^\sigma(\xp-\xpp) J^\tau(\xpp)\bigg\}\, .
\label{gen1}
\ee
We determine $\N$, which depends only on $J$, by inserting
Eq.~(\ref{gen1}) into Eq.~(\ref{dsi1}) or (\ref{qeqaf}):
\be
\ln\N(J)&=&\frac{i}{2}\int\dx\dxp J_{\mu}(x)D^{\mu\nu}(x-\xp)J_{\nu}(\xp),
\ee
resulting in the generating functional for the photonic sector
\be
\langle 0_+\left|\right. 0_-\rangle^{(J,\Jg)}_{0}
&=&\exp \bigg\{\frac{i}{2}
\int (dx)(dx') J_\mu(x)D^{\mu\nu}(x-x^\prime)J_\nu(x^\prime)\nonumber\\
&& \qquad\mbox{}+\frac{i}{2}\int(dx)(dx'){\Jg}_{\mu}(x)D^{\mu\nu}(x-x^\prime)
{\Jg}_{\nu}(x^\prime)
\nn
&&\qquad\mbox{}-i\int(dx)(dx')J_\mu(x)\tD^{\mu\nu}(x-\xp){\Jg}_\nu(\xpp)
\bigg\}\, ,
\label{phot}
\ee
where we use the shorthand notation for the ``dual
propagator'' that couples magnetic to electric charge
\be
\tD_{\mu\nu}\left(x-x'\right)=\epsilon_{\mu\nu\sigma\tau}
\int (dx'')
D_+\left(x-x''\right)
\partial^{\prime\prime\sigma} f^\tau\left(x''-x'\right).
\label{dualp}
\ee
The term coupling electric and magnetic sources has the same form as in
Eq.~(\ref{interact}); here,  we have replaced
$D^{\kappa\mu}\to g^{\kappa\mu}D_+$, as
we may because of the appearance of the Levi-Civita symbol in Eq.~(\ref{dualp}).
In an even more straightforward manner Eqs.~(\ref{dsi3}), (\ref{dsi4}) 
integrate to
\be
\langle 0_+\left|\right. 0_-\rangle^{(\bar{\eta},\eta,\bar{\xi},\xi)}_0=
\exp \left\{i\int (dx)(dx')
\left[\bar\eta(x) G_{\psi}(x -x')\eta(x') + \bar\xi(x) 
G_{\chi}(x -x')\xi(x')\right]\right\}\, ,
\label{fermi}
\ee
where $G_{\psi}$ and $G_{\chi}$ are the free propagators for the
electrically and magnetically charged fermions, respectively,
\be
G_{\psi}(x)&=&\frac{1}{-i\dslash +m_\psi}\delta(x)\, ,
\nn
G_{\chi}(x)&=&\frac{1}{-i\dslash +m_\chi}\delta(x)\, .
\ee
In the presence of interactions the coupled 
equations~(\ref{dsi1}), (\ref{dsi2}), (\ref{dsi3}), (\ref{dsi4}) are 
solved by substituting Eqs.~(\ref{phot}) and (\ref{fermi})
into Eq.~(\ref{vacint}). 
The resulting generating function is
\be
Z(\J)
&=&\exp\left(-ie\int (dx)\frac{\delta}
{\delta\eta(x)}\gamma^{\mu}\frac{\delta}{i\delta J^\mu(x)}
\frac{\delta}{\delta\bar\eta(x)}\right)
\exp\left(-ig\int(dy)\frac{\delta}{\delta\xi(y)}\gamma^{\mu}
\frac{\delta}{i\delta\Jg^\mu(y)}\frac{\delta}{\delta\bar\xi(y)}\right)
Z_0(\J)\, .
\label{genfunct}
\ee

\subsection{Nonperturbative Generating Functional}
\label{ss:np}

Due to the fact that any expansion in $\alpha_g$ or $eg$
is not practically useful we recast the generating 
functional (\ref{genfunct}) into a functional 
form better suited for a nonperturbative calculation of the four-point
Green's function.

First we utilize the well-known Gaussian combinatoric 
relation \cite{sym54,fri90};  moving  
the exponentials containing the interaction vertices in
terms of functional derivatives with respect to fermion sources
past the free fermion propagators, we obtain (coordinate labels are now
suppressed)
\be
Z(\J)&=&
\exp\left\{i\int\bar\eta\left(G_{\psi}
\left[1-e\gamma\cdot\frac{\delta}{i\delta J}G_{\psi}\right]^{-1}\right)\eta
+\Tr\ln\left(1-e\gamma\cdot\frac{\delta}{i\delta J}G_{\psi}\right)\right\}
\nonumber \\
&&\times\exp\left\{i\int\bar\xi\left(G_{\chi}
\left[1-g\gamma\cdot\frac{\delta}{i\delta \Jg}G_{\chi}\right]^{-1}\right)\xi
+\Tr\ln\left(1-g\gamma\cdot\frac{\delta}{i\delta \Jg}G_{\chi}\right)\right\}
Z_0(J,{}^*J)\, .\nn
\label{functfull}
\ee
Now, we re-express Eq.~(\ref{phot}), the non-interacting
part of the generating functional of the photonic action, $Z_0(J,\Jg)$,
using a functional Fourier transform, 
\be
Z_{0}(J,\Jg)=\int 
\left[d A\right]\left[d B\right]\tilde{Z_{0}}\left(A,B\right)
\exp\left[i\int\left(J\cdot A +
\Jg\cdot B\right)\right]\, ,
\label{piphot1}
\ee
or
\be
Z_0(J,\Jg)
=\int\, \left[d A\right]\left[d B\right] 
\exp\left(i\Gamma_0[A,B,J,\Jg]\right),
\ee
where (using a matrix notation for integration over coordinates)
\be
\Gamma_0[A,B,J,\Jg]=
\int\left(J\cdot A +\Jg \cdot B\right)
-\frac{1}{2}\int A^{\mu} K_{\mu\nu} A^{\nu}
+\frac{1}{2}\int 
 B^{\prime\mu}\tilde{\Delta}^{-1}_{\mu\nu}
B^{\prime\nu}
\label{onepi}
\ee
with the abbreviation
\be
B^{\prime}_\mu(x)=B_\mu(x)-\epsilon_{\mu\nu\sigma\tau}\int\dxp\partial^{\nu} 
f^{\sigma}(x-\xp) A^\tau(\xp)
\label{bprime}
\ee
and the string-dependent ``correlator''
\be
\tilde{\Delta}_{\mu\nu}(x-\xp)=\int \dxpp
\left\{f^{\sigma}(x-\xpp)f_{\sigma}\left(\xpp-\xp\right)g_{\mu\nu}
-f_{\mu}\left(x-\xpp\right)f_{\nu}\left(\xpp-\xp\right)\right\}
\label{strprp}
\ee
(see Appendix~\ref{a:pi} for details).  
Using Eq.~(\ref{onepi}) we recast Eq.~(\ref{functfull}) as
\be
Z(\J)&=&
\int\, \left[d A\right]\left[d B\right]
 F_{1}(A)F_{2}(B)\exp\left(i\Gamma_0[A,B,J,\Jg]\right)\, .
\ee
Here the fermion functionals $F_1$ and $F_2$ are obtained by the replacements
${\delta\over\i \delta J}\to A$, ${\delta\over i\delta{}\Jg}\to B$:
\be
F_{1}(A)&=&\exp\left\{\Tr\ln\left(1-e\gamma\cdot A G_{\psi}\right)+
i\int\bar\eta\left(G_{\psi}
\left[1-e\gamma\cdot A G_{\psi}\right]^{-1}\right)\eta\right\}\, ,
\nn
F_{2}(B)&=&\exp\left\{\Tr\ln\left(1-g\gamma\cdot B G_{\chi}\right) +
i\int\bar\xi\left(G_{\chi}
\left[1-g\gamma\cdot B G_{\chi}\right]^{-1}\right)\xi\right\}\, .
\label{loops}
\ee
We perform a change of variables by shifting about the stationary
configuration of the effective action, $\Gamma_0[A,B,J,\Jg]$:
\be
A_\mu(x)=\bar{A}_\mu(x)+\phi_\mu(x)\, ,
\hspace{1cm}
B'_\mu(x)=\bar{B}'_\mu(x)+\phi^\prime_\mu(x)
\label{exp}
\ee
where $\bar{A}$ and $\bar{B}$ are given by the solutions to
\be
\frac{\delta\Gamma_0\left(A,B,J,\Jg\right)}{\delta A^{\tau}}=0\, ,
\hspace{1cm}
\frac{\delta\Gamma_0\left(A,B,J,\Jg\right)}{\delta B^{\tau}}=0\, ,
\ee
namely (most easily seen by regarding $A$ and $B'$ as independent variables),
\be
\bar{A}_\mu(x)&=&\int\dxp D_{\mu\kappa}(x-\xp)\bigg(J^\kappa(\xp)
-\epsilon^{\kappa\nu\sigma\tau}\int\dxpp\partial^{\prime}_\nu
f_\sigma(\xp-\xpp)\Jg_\tau(\xpp)\bigg),
\nn
\bar{B}_\mu(x)&=&\int\dxp D_{\mu\kappa}(x-\xp)\bigg(\Jg^\kappa(\xp)
+\epsilon^{\kappa\nu\sigma\tau}\int\dxpp\partial^{\prime}_\nu
f_\sigma(\xp-\xpp)J_\tau(\xpp)\bigg),
\label{bgsol}
\ee
reflecting the form of Eq.~(\ref{inta}) and its dual.
Note that the solutions (\ref{bgsol}) respect the dual symmetry, which is not
however manifest in the form of the effective action (\ref{onepi}).
Using the properties of Volterra expansions for functionals
and performing the resulting quadratic integration over 
$\phi(x)$ and $\phi^\prime(x)$ (see Appendix B),
we obtain a rearrangement of the generating functional 
for the monopole-electron system that is well suited for 
nonperturbative calculations:
\be
\frac{Z(\J)}{Z_0(J,\Jg)}&=&
\exp\Bigg\{\frac{i}{2}\int(dx)(dx')
\Bigg(\frac{\delta}{\delta\bar{A}_{\mu}(x)}
 D_{\mu\nu}(x-x')\frac{\delta}{\delta\bar{A}_{\nu}(x')}
+\frac{\delta}{\delta\bar{B}_{\mu}(x)}
 D_{\mu\nu}(x-x')\frac{\delta}{\delta\bar{B}_{\nu}(x')}\Bigg)
\nn &&
\hspace{4cm}\mbox{}
-i\int(dx)(dx')\frac{\delta}{\delta\bar{A}_{\mu}(x)}
\tD_{\mu\nu}\left(x-\xp\right)
\frac{\delta}{\delta\bar{B}_{\nu}(\xp)}\Bigg\}
\nn
&& \times\exp\bigg\{ i\int(dx)(dx')\bar{\eta}\left(x\right)
G(x,\xp|\bar{A})\eta\left(\xp\right) 
+i\int(dx)(dx')\bar{\xi}\left(x\right)
G(x,\xp|\bar{B})\xi\left(\xp\right)\bigg\}
\nn
&& \times \exp\bigg\{-\int_{0}^{e} de^{\prime} \,
\Tr\gamma\bar{A} G(x,x|\bar{A})
-\int_{0}^{g} dg^{\prime}\, \Tr\gamma\bar{B}
G(x,x|\bar{B})\bigg\}\, .
\label{master}
\ee
Here the two-point fermion Green's functions $G(x_1,y_1|\bar{A})$, and 
$G(x_2,y_2|\bar{B})$ in the background of the stationary
photon field $\bar{A},\bar{B}$ are given by
\be
G(x,x'|\bar{A})&=&\langle x|
(\gamma p +m_\psi-e\, \bAslash)^{-1}| x'\rangle \, ,
\nn
G(x,x'|\bar{B})&=&\langle x|
(\gamma p +m_\chi-g\, \bBslash)^{-1}|x'\rangle\, ,
\ee
where the trace includes integration over spacetime.
This result is equivalent to the functional
Fourier transform given in Eq.~(\ref{piphot1}) including the fermionic
monopole-electon system:
\be
Z({\cal J})&=&\int
\left[d A\right]\left[d B\right] 
\det\left(-i\gamma D_A+m_\psi\right) \det\left(-i\gamma D_B+m_\chi\right)
\nn 
&&\times
\exp\Bigg\{ i\int(dx)(dx')\bigg(\bar{\eta}\left(x\right)
G(x,\xp|A)\eta\left(\xp\right) + \bar{\xi}\left(x\right)
G(x,\xp|B)\xi\left(\xp\right)\bigg)\Bigg\}
\nn
&&\times
\exp\left\{-\frac{i}{2}\int\left( A^{\mu} K_{\mu\nu} A^{\nu}-
 B^{\prime\mu}\tilde{\Delta}^{-1}_{\mu\nu}B^{\prime\nu}\right)
+ i\int\left(J\cdot A +\Jg \cdot B\right)\right\},
\label{world}
\ee
where we have integrated over the fermion degrees of freedom.

Finally from our knowledge of the manner in which electric
and magnetic charge couple to photons through Maxwell's equations
we can immediately write the generalization of Eq.~(\ref{master})
for dyons, the different species of which are labeled by the index $a$:
\be
Z(\J)&=&
\exp\Bigg\{\frac{i}{2}
\int(dx)(dx')\K^\mu(x)\D_{\mu\nu}\left(x-x^\prime\right)
\K^\nu\left(x^\prime\right)\Bigg\}
\nn
&&\times
\exp\Bigg\{\frac{i}{2}\int(dx)(dx')
\frac{\delta}{\delta\bar{\A}_{\mu}(x)}
\D_{\mu\nu}(x-x')\frac{\delta}{\delta\bar{\A}_{\nu}(x')}\Bigg\}
\nn
&&\times
\exp\bigg\{ i\sum_a\int(dx)(dx')\bar{\zeta}_a\left(x\right)
G_a(x,\xp|\bar{\A_a})\zeta_a\left(\xp\right)\bigg\}
\nn
&&\times 
\exp\bigg\{-\sum_a\int_{0}^{1} dq \,
\Tr\gamma\bar{\A_a} G_a(x,x|q\bar{\A_a})\bigg\}\, .
\label{masterd}
\ee
where $\A_a=e_aA+g_aB$, $\zeta_a$ is the source for the dyon of species $a$,
and a matrix notation is adopted,
\begin{mathletters}
\be
\K^{\mu}(x)=
\left(
\begin{array}{c}
J(x)\\ \Jg(x)
\end{array}
\right),
\hspace{.5cm}
\frac{\delta}{\delta\bar{\A}_{\mu}(x)}=
\left(
\begin{array}{c} 
\delta/\delta\bar{A}_{\mu}(x) \\
\delta/\delta\bar{B}_{\mu}(x)
\end{array} 
\right),
\ee
\end{mathletters}
and
\begin{mathletters}
\be
\D_{\mu\nu}\left(x-x^\prime\right)&=&
\left(
\begin{array}{cc}
D_{\mu\nu}\left(x-x^\prime\right) & -\tD_{\mu\nu}\left(x-x^\prime\right) \\
\tD_{\mu\nu}\left(x-x^\prime\right) & D_{\mu\nu}\left(x-x^\prime\right)
\end{array}
\right)\, .
\ee
\end{mathletters}

\section{String Independence of the Dyon-Dyon  Scattering
Cross Section} 

In this section we demonstrate  the string independence
of the dyon-dyon and charge-monopole (the latter being a
special case of the former) scattering cross section.  We will use
the generating functional (\ref{masterd}) developed in the last 
section to calculate the  scattering cross section
nonperturbatively.  We find that we are able to demonstrate 
phenomenological string invariance of the scattering cross section. 
It appears that in much the same manner as the Coulomb phase
arises as a soft effect in high energy charge scattering,
the string dependence arises from the exchange of soft
photons.  

To calculate the dyon-dyon scattering cross section
we obtain the four--point Green's
function for this process from Eq. (\ref{masterd})
\be
G\left(x_1,y_1;x_2,y_2\right)&=&
\frac{\delta}{i\delta\bar{\zeta}_1(x_1)}
\frac{\delta}{i\delta{\zeta}_1(y_1)}
\frac{\delta}{i\delta\bar{\zeta}_2(x_2)}
\frac{\delta}{i\delta{\zeta}_2(x_2)}
Z(\J)\bigg|_{{\cal J}=0}.
\label{gnfnct3}
\ee  
The subscripts on the sources refer to the two different dyons.

\begin{figure}
\centerline{
\epsfig{figure=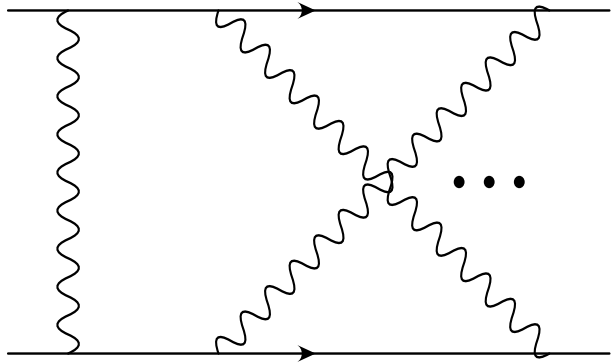,height=5cm,width=8cm}}
\caption{Dyon-dyon scattering amplitudes  in the quenched approximation.}
\label{figlink} 
\end{figure}

Here we confront our calculational limits; these are not too
dissimilar from those encountered in diffractive scattering  or 
in the strong-coupling regime
of QCD~\cite{nac91,kor94,gr98,gr98a}.
As a first step in analyzing the string dependence
of the scattering amplitudes,
we study high-energy forward scattering
processes where {\em soft} photon contributions
dominate.
In diagrammatic language, in this kinematic regime 
it is customary to restrict attention to 
that subclass in which there are no closed fermion loops
and the photons are exchanged 
between fermions~\cite{nac91}.  In the context of Schwinger-Dyson 
equations this amounts to quenched or ladder
approximation (see Fig.~\ref{figlink}).
In this approximation
the linkage operators, $\L$,
connect two fermion propagators 
via photon exchange, as we read off from
Eq.~(\ref{master}):
\be
\e^{\L_{12}} =\exp\Bigg\{i
\int \dx\dxp\frac{\delta}{\delta\bar{\A}^{\mu}_1(x)}
\D^{\mu\nu}\left(x-\xp\right)
\frac{\delta}{\delta \bar{\A}^{\nu}_2(\xp)}\Bigg\}\, .
\label{link}
\ee
In this approximation Eq.~(\ref{gnfnct3}) takes the form
\be
G\left(x_1,y_1;x_2,y_2\right)&=&
-\e^{\L_{12}}
G_1(x_1,y_1|\bar\A_1)G_2(x_2,y_2|\bar\A_2)\Big|_{\bar{A}=\bar{B}=0}\, ,
\label{scatt}
\ee
where we express the two-point function using the proper-time 
parameter representation of an ordered exponential
\be
G_a(x,y|\bar{\A}_a)&=&i\int_{0}^{\infty}d\xi\,
\e^{-i\xi\left(m_{a}-i\gamma\partial\right)}
\exp\left\{{i\int_{0}^{\xi}d\xi^{\prime}{\rm e}^{\xi^{\prime}
\gamma\partial}\gamma\bar\A_a{\rm e}^{-\xi^{\prime}
\gamma\partial}}\right\}_{+}\delta(x-y)
\label{gextful}
\ee
where ``$+$'' denotes path ordering in $\xi^{\prime}$.
The 12 subscripts in $\L_{12}$ emphasize that only photon lines that
link the two fermion lines are being considered.

\subsection{High Energy Scattering Cross Section}
\label{s:ea}

Adapting techniques outlined in \cite{qui76,fri97} we consider 
the connected form of Eq.~(\ref{scatt}).
We use the connected two-point function and the identities
\be
\e^{\L}=1+\int_0^1 da\, \e^{a\L}\L
\label{par}
\ee
and 
\be
\frac{\delta}{\delta\bA_{\mu}(x)} G(y,z|\bar{\A})
=e G(y,x|\bar{\A})\gamma^\mu
G(x,z|\bar{\A})\, .
\label{gderiv}
\ee
Using Eqs.~(\ref{scatt}) and (\ref{gextful})
one straightforwardly  is led to the
following representation of the four-point
Green function,
\be
&&G(x_{1},y_{1};x_{2},y_{2})=
-i\int_0^1 da\int \dzi\dzii\left(\cq\,
D_{\mu\nu}(z_1-z_2)
-\wq\tD_{\mu\nu}(z_1-z_2\right))
\nn
&& 
\quad\times\e^{a\L_{12}}
G_1(x_{1},z_1|\bar{\A}_1)\gamma_\mu
G_1(z_1,y_{1}|\bar{\A}_1)
G_2(x_{2},z_2|\bar{\A}_2)\gamma^\nu
G_2(z_2,y_{2}|\bar{\A}_2)\Bigg|_{\bar A=\bar B=0}\, ,
\label{mxptl}
\ee 
where the charge combinations  invariant under duality transformations are 
\be
\cq&=&e_1e_2+g_1g_2
\nn
\wq&=&e_1g_2-g_1e_2\, .
\label{chginv}
\ee

In order to account for  the soft nonperturbative effects of
the interaction between electric and magnetic charges
we consider the limit in which the momentum exchanged by the photons 
is small compared to the mass of the fermions. 
This affords a substantial simplification in evaluating
the path-ordered exponential in Eq.~(\ref{gextful});
in conjunction with the assumption of small
momentum transfer compared to the incident and outgoing
momenta, $q/p_{(1,2)}\ll\, 1$, this amounts to 
the Bloch-Nordsieck~\cite{blo37} or {\it eikonal approximation}
(see~\cite{fri65,lev69,ab69,dit70} and for more modern
applications in diffractive and strong coupling QCD processes 
\cite{nac91,kor94,gr98,gr98a}). 
In this approximation Eq.~(\ref{gextful}) becomes 
\be
G_a(x,y|\bar{\A})&=&
i\int_{0}^{\infty}d\xi\,
\e^{-i\xi m}\delta\left(x-y-\xi\frac{p}{m}\right)
\exp\left\{i\int_{0}^{\xi}d\xi^{\prime}
\frac{p}{m}\cdot\bar{\A}\left(x-\xi'\frac{p}{m}\right)
\right\}\, .
\label{gext}
\ee
With this simplification each propagator 
in Eq.~(\ref{scatt}) can be written
as an exponential of linear function of the gauge field.
Performing mass shell amputation on each external coordinate
and taking the Fourier transform of Eq.~(\ref{mxptl}) we obtain
the scattering 
amplitude,  $T(p_{1},p^{\prime}_{1};p_{2},p^{\prime}_{2})$:
\be
T\left(p_1,\pp_1 ; p_2,\pp_2\right)
&=& 
-i\int_0^1 da\, 
\e^{a\L_{12}}
\int \dzi\dzii\left(\cq\,
D_{\mu\nu}\left(z_1-z_2\right)
-\wq\tD_{\mu\nu}\left(z_1-z_2\right)\right)
\nn
&&\hspace{-2cm}\times
\int\dxi\e^{-ip_1x_1}
\bar{u}(p_1)\left(m_1+v_1\cdot p_1\right)
G_1(x_1,z_1|\bar{\A}_1)\gamma^\mu
\int\dyi\e^{i\ppi y_1}
G_1(z_1,y_1|\bar{\A}_1)
\left(m_1+v_1^\prime\cdot\ppi\right)u(\ppi)
\nn
&&
\hspace{-2cm}\times
\int\dxii\e^{-ix_2 p_2}
\bar{u}(p_2)\left(m_2+v_2\cdot p_2\right)
G_2(x_2,z_2|\bar{\A}_2)\gamma^\nu
\int\dyii\e^{i\ppii y_2}
G_2(z_2,y_2|\bar{\A}_2)
\left(m_2+v_2^\prime\cdot\ppii\right)u(\ppii).
\label{4.10}
\ee
Substituting Eq.~(\ref{gext}) into Eq.~(\ref{4.10}), we simplify this to
\be
T(p_{1},p^{\prime}_{1};p_{2},p^{\prime}_{2})&=&-i\int_0^1 d\, a\, 
\int\dzi\dzii
\e^{-iz_{1}\left(p_{1}-p^{\prime}_{1}\right)}
\e^{-iz_{2}\left(p_{2}-p^{\prime}_{2}\right)}
\bar{u}(p^{\prime}_{1})\gamma^{\mu}u(p_{1})
\bar{u}(p^{\prime}_{2})\gamma^{\nu}u(p_{2})
\nn
&&
\times\left(\cq\,
D_{\mu\nu}(z_{1}-z_{2})-\wq\tD_{\mu\nu}(z_{1}-z_{2})\right)
\e^{a\L_{12}}
\nn
&&\times
\exp\left[i\int_{0}^{\infty}d\alpha_{1}\left\{
p_{1}\cdot\left(\bar{\A}_1\left(z_{1}+\alpha_{1} p_{1}\right)\right)
+p^{\prime}_{1}\cdot
\left(\bar{\A}_1\left(z_{1}-\alpha_{1} p^{\prime}_{1}\right)
\right)\right\}\right]\nn
&&\times\exp\left[
i\int_{0}^{\infty}d\alpha_{2}
\left\{
p_{2}\cdot\left(\bar{\A}_2\left(z_{2}+\alpha_{2} p_{2}\right)\right)
+p^{\prime}_{2}\cdot
\left(\bar{\A}_2\left(z_{2}-\alpha_{2} p^{\prime}_{2}\right)\right)
\right\}\right] \, .
\label{smp}
\ee

Choosing the incoming momenta to be in the $z$
direction, in the center of momentum frame,
 $p_1^\mu=(E_1,0,0,p)$, $p_2^\mu=(E_2,0,0,-p)$,
invoking the approximation of small recoil 
and passing  the linkage operator through the exponentials containing
the photon field, we find from Eq.~(\ref{smp})
\be
T(p_{1},p^{\prime}_{1};p_{2},p^{\prime}_{2})&=&-i\int_0^1 da 
\int\dzi\dzii
\e^{-iz_{1}\left(p_{1}-p^{\prime}_{1}\right)}
\e^{-iz_{2}\left(p_{2}-p^{\prime}_{2}\right)}
\bar{u}(p^{\prime}_{1})\gamma_{\mu}u(p_{1})
\bar{\upsilon}(p^{\prime}_{2})\gamma_{\nu}\upsilon(p_{2})
\nn 
&&
\times\left(\cq\,
D^{\mu\nu}\left(z_{1}-z_{2}\right)
-\wq\tD^{\mu\nu}\left(z_{1}-z_{2}\right)\right)
\e^{ia\Phi\left(p_1,p_2;z_1-z_2\right)}\, ,
\label{smp2}
\ee
where the ``eikonal phase'' integral is,
\be
\Phi_n\left(p_1,p_2;z_1-z_2\right)=
p_{1}^{\kappa}p_{2}^{\lambda}
\int_{-\infty}^{\infty}d\alpha_{1}\, d\alpha_2
\left(\cq\,
D_{\kappa\lambda}
 -\wq\tilde D_{\kappa\lambda}\right)
 \left(z_{1}-z_{2}+\alpha_1 p_{1}-\alpha_2 p_{2}\right)\, .
\label{phase1}
\ee
We transform to the center of momentum coordinates, by decomposing the
relative coordinate accordingly,
\be
\left(z_1-z_2\right)^\mu
=x_{\bot}^\mu-\tau_1 p_1^\mu+\tau_2 p_2^\mu,
\ee
where the Jacobian of the transformation is
\be
J=p\sqrt{s}
\ee
and $s=-(p_1+p_2)^2$ is the square of the center of mass energy. 
Here we use the {\em symmetric} (see~\cite{mil76,mil78} for
details) infinite string function, which has the momentum-space form,
\be
f^\mu(k)=\frac{n^\mu}{2\, i}\left(\frac{1}{n\cdot k-i\delta}+
\frac{1}{n\cdot k+i\delta}\right).
\label{infstring}
\ee
Inserting the momentum-space representation of the propagator, we cast
 Eq. (\ref{phase1}) into the form
\be
\Phi_n\left(p_1,p_2;x\right)&=& 
p_{1}^{\kappa}p_{2}^{\lambda}
\int_{-\infty}^{\infty}d\alpha_1\, d\alpha_2
 \int\frac{(dk)}{\left(2\pi\right)^{4}}
\frac{\e^{i\, k\cdot\left(x+\alpha_1 p_{1}-\alpha_2 p_{2}\right)}}
{ k^{2}+\mu^2}
\nn
&& 
\times\left[\cq\, g_{\kappa\lambda} -
\wq\epsilon_{\kappa\lambda\sigma\tau}k^{\sigma} 
\frac{n^{\tau}}{2}
\left(\frac{1}{n\cdot k-i\delta}+\frac{1}{n\cdot k+i\delta}\right)\right],
\label{phase2}
\ee
where we have introduced the standard infrared photon-mass regulator,
$\mu^2$.
The delta functions that result from performing the integrations 
over parameters $\alpha_1$ and $\alpha_2$ in Eq.~(\ref{phase2})
in the eikonal phase suggests the momentum decomposition
\be
k^\mu &=& k_{\perp}^\mu + \lambda_1 e_1^\mu + \lambda_2 e_2^\mu\, ,\,\, 
\text{where}\,\,  
\lambda_1 = p_2\cdot k\, ,\, \text{and}\,\, 
\lambda_2 = p_1\cdot k ,
\ee
and the four-vector basis is given by
\be
e_1^\mu=\frac{-1}{\sqrt{s}}\left(1,0,0,\frac{p_1^0}{p}\right)\quad
\text{and}\quad 
e_2^\mu=\frac{-1}{\sqrt{s}}\left(1,0,0,-\frac{p_2^0}{p}\right)\, ,
\ee
which have the following properties,
\be
e_1\cdot e_1&=&\frac{1}{s}\frac{M_1^2}{ p^2}, \quad 
e_2\cdot e_2=\frac{1}{s}\frac{M_2^2}{ p^2}, 
\quad\text{and}\quad  
e_1\cdot e_2 =\frac{1}{s}\frac{p_1\cdot p_2}{ p^2}\, .
\ee
The corresponding measure and Jacobian are, respectively,
\be
\left(dk\right)&=&Jd^2\bbox{k}_\perp d\lambda_1
d\lambda_2\quad\mbox{and}\quad J=\left(p\sqrt{s}\right)^{-1}\, .
\ee
Using the
definition of the M{\o}ller amplitude, $M(s,t)$, given by removing the
momentum-conserving delta function,
\be
T(p_{1},p^{\prime}_{1};p_{2},p^{\prime}_{2})
=(2\pi)^4\delta^{(4)}\left(P-P^{\prime}\right)M(s,t)\, ,
\ee
we put Eq.~(\ref{smp2}) into the form
\be
M(s,t)=
-i\int_0^1 da\int  d^2\bbox{x}_{\bot}
\e^{-i\bbox{q}_{\bot}\cdot\bbox{x}_{\bot}}
\bar{u}(p^{\prime}_{1})\gamma^{\mu}u(p_{1})
\bar{u}(p^{\prime}_{2})\gamma^{\nu}u(p_{2}) I_{\mu\nu}
\e^{ia\Phi_{n}\left(p_1,p_2 ; x\right)}\, ,
\label{smp3}
\ee
where 
\be
I_{\mu\nu}&=&\int\frac{d^2\bbox{k}_{\bot}}{(2\pi)^2}
\frac{d\lambda_{1}}{2\pi}\frac{d\lambda_{2}}{2\pi}
\frac{\e^{i\bbox{k}_{\bot}\cdot\bbox{x}_{\bot}}
2\pi\delta(\lambda_1)2\pi\delta(\lambda_2)}
{\left(\bbox{k}_{\bot}^2+\mu^2+ \frac{1}{s\, p^2}
\left(\lambda_1^2 M_1^2+\lambda_2^2 M_2^2
+2\lambda_1\lambda_2 p_1\cdot p_2\right)\right)}
\nn
&&\hspace{1cm}
\times\left[\cq\, g_{\mu\nu}
-\wq\epsilon_{\mu\nu\sigma\tau}
k^{\sigma}\frac{n^{\tau}}{2}
{\left(\frac{1}{n\cdot k-i\delta}+\frac{1}{n\cdot k+i\delta}\right)}\right]
\ee
Here $P=p_1+p_2$ and $P^\prime=p_1^\prime+p_2^\prime$,
and $q=p_1-p_1'$ is the momentum transfer. The factor 
\be
\exp(i\tau_1p_1\cdot q-i\tau_2p_2\cdot q)=
\exp\left[i{1\over2}q^2(\tau_1+\tau_2)\right]
\ee
has been omitted because it is unity in the eikonal limit, and correspondingly,
we have carried out the integrals on $\tau_1$ and $\tau_2$.
The eikonal phase Eq.~(\ref{phase2}) now takes the very similar form
\be
\Phi_{n}\left(p_1,p_2;x\right)=
\frac{p_{1}^{\kappa} p_{2}^{\lambda}}{p\sqrt{s}}I_{\kappa\lambda}.
\label{phase4}
\ee
Choosing a spacelike 
string\footnote{We choose a spacelike string in order
that we formally have a local interaction in momentum space.}, 
$n^\mu=(0,\bbox{n})$, integrating over 
the coordinates $\lambda_1 , \lambda_2$,
and introducing ``proper-time'' parameter representations
of the propagators, we reduce Eq.~(\ref{phase4}) to
\be
\Phi_{n}\left( p_1, p_2;x\right)&=&{1\over p\sqrt{s}}
\int \frac{d^2\bbox{k}}{(2\pi)^2}\e^{i\bbox{k}\cdot\bbox{x}}
\int_{0}^{\infty} d\ps\,\e^{-\ps\left(\bbox{k}^2+\mu^2\right)}
\nn
&& \hspace{1cm}
\times\Bigg\{\cq\,p_1\cdot p_2  -
\wq p_1^\mu p_2^\nu\epsilon_{\mu\nu\sigma\tau}\frac{n^\sigma}{2\, i}
\frac{\partial}{\partial n_\tau}
\left(
\int_{0}^{\infty}\,\frac{d\pt}{i\pt}\,
\e^{i\pt\left(\bbox{n}\cdot\bbox{k}+i\delta\right)}
-\int_{-\infty}^{0}\,\frac{d\pt}{i\pt}\,
\e^{i\pt\left(\bbox{n}\cdot\bbox{k}-i\delta\right)}\right)
\Bigg\}
\nn
&=&\frac{1}{2\pi}\left\{
\cq\, \frac{p_1\cdot p_2}{p\sqrt{s}} 
K_0\left(\mu\left|\bbox{x}\right|\right)-
\wq\epsilon_{3jk} n^j\frac{\partial}{\partial n^k}
{1\over2}\int\frac{d\pt}{\pt}\,
K_0\left(\mu\left|\left(\bbox{x}+\pt\bbox{n}\right)\right|\right)\right\},
\label{phase5}
\ee
in terms of modified Bessel functions,
where we have dropped the subscript ${\small \bb{\perp}}$.

We perform the parameter integral over $\pt$
in the limit of small $\mu^2$:
\be
-{1\over 2}\bbox{\hat z\cdot(\hat n\times x)}
\left[\int_0^\infty-\int_{-\infty}^0
\right]{d\pt\,e^{-\delta |t|}\over(\pt+\bbox{\hat n\cdot x})^2+x^2-(
\bbox{\hat n\cdot
x})^2}=\arctan\left[{\bbox{ n\cdot x}\over
\bbox{\hat z\cdot(\hat n\times x)}}\right],
\ee
so the phase is
\be
\Phi_{n}\left( p_1, p_2;x\right)=\frac{1}{2\pi}\left\{\cq
\ln\left(\tilde{\mu}\left|\bbox{x}\right|\right)
-\wq\arctan\left[\frac{\bbox{\hat n}\cdot\bbox{x}}
{\bbox{\hat z}\cdot\left(\bbox{\hat n}\times\bbox{x}\right)}\right]
\right\}.
\ee
In this limit we have used the asymptotic limit of
the modfied Bessel function
\be
K_0(x)\sim -\ln\left(\frac{\e^\gamma x}{2}\right)\, ,
\ee
where $\gamma=0.577\dots$ is Euler's constant
and we have defined
$\tmu=\e^\gamma\mu /\,2$.  
Similarly,  Eq.~(\ref{smp3}) becomes
\be
M(s,t)&=&
-{i\over2\pi}\int_0^1 da\int d^2\bbox{x}\,\e^{i\bbox{q}\cdot\bbox{x}}
\bar{u}(p^{\prime}_{1})\gamma^{\mu}u(p_{1})
\bar{u}(p^{\prime}_{2})\gamma^{\nu}u(p_{2})
\nn
&&
\times\Bigg\{
g_{\mu\nu}\cq K_0\left(\mu\left|\bbox{x}\right|\right)
-\epsilon_{\mu\nu\sigma\tau}\wq
n^\tau\frac{\partial}{\partial n_\sigma}
{1\over2}\int\,\frac{d\pt}{\pt}\,
K_0
\left(\mu\left|\left(\bbox{x}+\pt\bbox{n}\right)\right|\right)\Bigg\}
\e^{ia\Phi_{n}\left(p_1,p_2 ; x\right)}.
\label{smp5}
\ee

Although in the eikonal limit, no spin-flip processes occur, it is,
as always, easier to calculate the helicity amplitudes, of which there
is only one in this case.  In the high-energy limit, $p^0\gg m$, the Dirac
spinor in the helicity basis is
\be
u^\sigma(p)=\sqrt{p^0\over 2m}(1+i\gamma_5\sigma)v_\sigma, 
\ee
where the $v_\sigma$ may be thought of as two-component spinors satisfying
$\gamma^0v_\sigma=v_\sigma$.  They are further eigenstates of the helicity
operator $\bbox{\sigma\cdot\hat p}$ with eigenvalue $\sigma$:
\be
v_+^\dagger(\bb{\hat{p}}^\prime)=\left( \begin{array}{c} 
\cos\frac{\theta}{2}, \sin\frac{\theta}{2}  
\end{array} \right)
\quad
v_-^\dagger(\bb{\hat{p}}^\prime)=
\left( 
\begin{array}{c} 
-\sin\frac{\theta}{2},\cos\frac{\theta}{2}  
\end{array} \right)
\quad
v_+(\bb{\hat{p}})=\left( 
\begin{array}{c} 1  \\  0
\end{array} \right)
\quad
v_-(\bb{\hat{p}})=\left( 
\begin{array}{c} 0  \\  1
\end{array} \right).
\ee
We employ the definition 
\be
\gamma_5=\gamma_0\gamma^1\gamma^2\gamma^3
\ee
and consequently $\gamma^0\bb{\gamma}=i\gamma^5\bb{\sigma}$, where
$\sigma_{ij}=\epsilon_{ijk}\sigma^k$. 
We then 
easily find upon integrating over the parameter $a$ that the spin non-flip
part of Eq.~(\ref{smp5}) becomes ($\theta\to0$)
\be
M(s,t)=\frac{s}{2M_1M_2}
\Bigg\{\int d^2\bbox{x}\, \e^{i\bbox{q}\cdot\bbox{x}}
\e^{i\Phi_{n}\left(p_1,p_2 ; x\right)}
-\left(2\pi\right)^2\delta^2\left(\bbox{q}\right)
\Bigg\}.
\label{scat9}
\ee

Now notice that the arctangent function is discontinuous
when the $xy$ component of $\bbox{\hat n}$
 and $\bbox{x}$ lie in the same direction.  We require that the eikonal phase
factor $e^{i\Phi_n}$ 
be continuous, which leads to the Schwinger quantization condition
 (\ref{dyon}):
 \be
\wq= 4N\pi\, .
\ee
Now using the integral form for 
the Bessel function of order $\nu$
\be
i^\nu\/J_\nu(t)=\int_0^{2\pi}\frac{d\phi}{2\pi}
\e^{i\left(t\cos\phi-\nu\phi\right)}\, ,
\ee
we find the dyon-dyon  scattering amplitude (\ref{scat9}) to be
\be
M(s,t)=\frac{\pi s}{M_1M_2}\e^{i2N\psi}
\int_0^\infty dx\,x \,J_{2N}\left(qx\right)
\e^{i2\tilde{\alpha}\ln\left(\tmu x\right)}\, ,
\label{scatbes}
\ee
where $\tilde{\alpha}=\cq/4\pi$, and $\psi$ is the angle between
$\bbox{q}$ and $\bbox{n}$.
The integral over $x$ is just a ratio of gamma functions,
\be
\int_0^\infty dx\,\left(\tmu x\right)^{1+2i\talpha}
 J_{2N}\left(qx\right)
=\frac{1}{2\tmu}
\left(\frac{4\tmu^2}{q^2}\right)^{i\talpha+1}
\frac{\Gamma\left(1+N+i\talpha\right)}
{\Gamma\left(N-i\talpha\right)}.
\label{infr}
\ee
Then Eq.~(\ref{scatbes}) becomes
\be
M(s,t)=\frac{ s}{M_1M_2}{2\pi\over q^2}(N-i\talpha)
\e^{i2N\psi}
\left(\frac{4\tmu^2}{q^2}\right)^{i\talpha}
\frac{\Gamma\left(1+N+i\talpha\right)}
{\Gamma\left(1+N-i\talpha\right)}.
\ee
This result is almost identical in structure to the nonrelativistic
form of the scattering amplitude for the Coulomb potential, which result
is recovered by setting $N=0$.  (See, for example, Ref.~\cite{gottfried}.)
Following the standard convention\cite{iz} we calculate
the spin-averaged cross section for dyon-dyon scattering
in the high energy limit,
\be
\frac{d\sigma}{dt}&=&
\frac{\left(\cq\right)^2+\left(\wq\right)^2}{4\pi t^2}
 \, .
\label{dyoncross}
\ee
While the Lagrangian is string-dependent, because of
the charge quantization condition, the cross section, Eq~(\ref{dyoncross}),
is string independent.

For the case of charge-monopole scattering $e_1=g_2=0$,
this result, of course, coincides with that
found by Urrutia \cite{urr78},
which  is  also  string independent as a consequence 
of (\ref{quan}).  This is to be contrasted with {\it ad hoc\/} prescriptions
that average over string directions  or eliminate its
dependence by simply dropping string-dependent
terms because they cannot contribute to any gauge invariant quantities
(\cf\ Ref.~\cite{dea82}).

\section{Conclusion}

In this paper we have responded to the challenge of Schwinger \cite{sch75},
to construct a realistic theory of relativistic magnetic charges.  He
sketched such a development in source theory language, but restricted his
consideration to classical point particles, explicitly leaving the details
to the reader.  Urrutia applied this skeletal formulation in the eikonal
limit \cite{urr78}, as already suggested by Schwinger.

We believe that we have given a complete formulation, in modern quantum
field theoretic language, of an interacting electron-monopole  or
dyon-dyon system.
The resulting Schwinger-Dyson equations, although to some extent implicit
in the work of Schwinger and others, are given here for the first time.

The challenge is to apply these equations to the calculation of monopole and dyon
processes.  Perturbation theory is useless, not only because of the strength
of the coupling, but more essentially because the graphs are fatally string-
(or gauge-) dependent.  The most obvious nonperturbative technique for 
transcending these limitations in scattering processes lies in the 
high energy regime where the 
eikonal approximation is applicable; in that limit, our formalism 
generalizes  the lowest-order result of Urrutia and
charts the way to include systematic corrections.  
More problematic is the treatment of monopole production 
processes---we defer that discussion to a subsequent publication.

In addition we have also detailed
how the Dirac string dependence disappears from physical quantities.
It is by no means a result of string 
averaging or a result of dropping
string-dependent terms as in~Ref.~\cite{dea82}.
In fact, it is a result of summing the soft contributions to 
the dyon-dyon or charge-monopole process.  There is good reason
to believe that inclusion of hard scattering contributions will
not spoil this consistency. At the level of the eikonal
approximation and its corrections
one might suspect the occurence of a 
factorization of hard string-independent
and soft string-dependent contributions in a manner similar to that
argued recently in strong-coupling QCD.

It is also of interest to
investigate other nonperturbative methods of calculation in order
to demonstrate gauge invariance of Green's functions and scattering 
amplitudes in both electron-monopole and dyon-dyon scattering
and in Drell-Yan production processes.\footnote{In addition there is a 
formalism recently employed in Ref.~\cite{fri95} based on 
Fradkin's \cite{fra66} Green's function representation, 
which includes approximate vertex and self-energy polarization 
corrections using nonperturbative techniques.} In a subsequent paper
we will apply the techniques and results found here to the Drell-Yan
production of monopole-antimonopole processes, and obtain phenomenologically
relevant estimates for the laboratory production of monopole-antimonopole
pairs.

\section*{Acknowledgements}

We are grateful to George Kalbfleisch for many useful discussions on
the physics of monopoles, and for reviving our interest in the whole
subject.  We further thank Markus Quandt for a careful reading of the
manuscript, and for useful suggestions, and Walter Dittrich for helpful
conversations.
This work is supported in part by a grant from the US
Department of Energy.

\appendix

\section{Path Integral}
\label{a:pi}

In this appendix we 
summarize the main steps to obtain the covariant path integral 
for the string dependent action corresponding to the generating
functional, Eq.~(\ref{genfunct}).  

This path integral is obtained by calculating the functional 
Fourier transform of  $Z_{0}$  
which in  the photonic sector  amounts to transforming Eq.
(\ref{phot}), according to Eq.~(\ref{piphot1}),
the functional transform of which is given by
\be
\hspace{-1.5cm}
\tilde{Z_{0}}\left(A,B\right)=
\int 
\left[d\, J\right]\left[d\Jg\right]Z_{0}(J,\Jg)\exp \left[-i\int 
\left(J\cdot A +\Jg\cdot B\right)\right]\, .
\label{c2}
\ee
After performing the Gaussian functional integration over $J$ in Eq.~(\ref{c2}), 
we obtain
\be
\tilde Z_0(A,B)&=&\int [d{}^*J]\exp\bigg\{-i\int (dx)\Jg_\mu(x)B^\mu(x)
\nonumber\\
&&\quad\mbox{}+{i\over2}\int(dx)(dx')\Jg_\mu(x)D^{\mu\nu}(x-x')\Jg_\nu(x')
\nonumber\\
&&\quad\mbox{}-{i\over2}\int(dx)(dx')A'_\mu(x)K^{\mu\nu}(x-x')A'_\nu(x')\bigg\},
\label{intoverj}
\ee
where $K_{\mu\nu}$ is the kernel given by Eq.~(\ref{ker}), the inverse
to $D^{\mu\nu}$, and 
\be
A^\prime_\mu(x)=A_\mu(x)+\epsilon_{\mu\nu\sigma\tau}\int(dx)(dx')(dx'')
D(x-x')\partial^{\prime \nu}f^\sigma(x'-x'')\Jg^\tau(x'').
\ee
We now use the following identity involving the contraction of $\epsilon$
symbols:
\be
\epsilon^{\mu\alpha\beta\gamma}\epsilon_{\mu\nu\sigma\tau}
&=&-g^\alpha_\nu g^\beta_\sigma g^\gamma_\tau
+g^\alpha_\nu g^\beta_\tau g^\gamma_\sigma
+g^\alpha_\sigma g^\beta_\nu g^\gamma_\tau
-g^\alpha_\sigma g^\beta_\tau g^\gamma_\nu\nn
&&\mbox{}-g^\alpha_\tau g^\beta_\nu g^\gamma_\sigma
+g^\alpha_\tau g^\beta_\sigma g^\gamma_\nu,
\ee
to simplify the final term in the exponential in Eq.~(\ref{intoverj}):
\be
&&-{i\over2}\int(dx)(dx') A'_\mu(x) D^{-1}(x-x') A^{\prime\mu}(x')
=-{i\over2}\int(dx)(dx') A_\mu(x) D^{-1}(x-x')A^{\mu}(x')\nn
&&\mbox{}+i\int(dx)(dx') \Jg^\mu(x)\epsilon_{\mu\nu\sigma\tau}\partial^\nu
f^\sigma(x-x')A^\tau(x')
-{i\over2}\int(dx)(dx')\Jg_\mu(x)\Delta^{\mu\nu}(x-x')
\Jg_\nu(x')\nn
&&\mbox{}-{i\over2}\int(dx)(dx')\Jg^\mu(x)D_+(x-x')\Jg_\mu(x'),
\label{a5}
\ee
the last term in which cancels the second term in the exponential in
Eq.~(\ref{intoverj}).  Here we see the ``string propagator,''
 Eq.~(\ref{strprp}). Now we carry out the $\Jg$ functional integration,
 noticing that the second term on the right side of Eq.~(\ref{a5})
 converts $B^\mu$ to $B^{\prime\mu}$, Eq.~(\ref{bprime}):
\be
Z_{0}(A,B)&=&\int\left[d A\right]\left[d B\right]
\exp\bigg\{-\frac{i}{2}\int(dx)(dx') A^\mu(x) K_{\mu\nu}(x-\xp) A^\nu(\xp)
\nonumber \\
&&\qquad\mbox{}+\frac{i}{2}\int(dx)(dx')B^{\prime\mu}(x)
\tilde{\Delta}^{-1}_{\mu\nu}(x-x^\prime)B^{\prime\nu}(x^\prime)\bigg\}\,
\label{twopot}
\ee
which implies the effective action (\ref{onepi}).

\section{Functional Rearrangement}
\label{a:fnct}

We consider the expansion about the minima of the effective action 
$\Gamma_0[A,B,J,\Jg]$,
in particular, the impact on Eqs.~(\ref{loops}) of
the transformation~(\ref{exp}).
Using the properties of Volterra expansions for functionals,
the shift in variables results in the translation of the 
loop  functionals
\begin{mathletters}
\be
&&F_1\left(\bar{A}_\mu+\phi_\mu\right)=
\exp\left\{i\int(dx) \phi_\alpha(x)\frac{\delta}{i\delta\bar{A}_\alpha(x)}\right\}
F_1\left(\bar{A}_\mu\right)\, , 
\label{f1}
\\
&&F_2\bigg(\hat{B}^\prime_\mu+\phi_\mu^\prime\bigg)
=\exp\Bigg\{i\int(dx)\bigg(\phi_\alpha^{\prime}(x)-
\epsilon_{\alpha\beta\gamma\delta}
\int\dxp \partial^\beta f^\gamma(x -\xp)\phi^\delta(\xp)
\bigg)\frac{\delta}{i\delta\bar{B}_\alpha(x)}\Bigg\}F_2\left(\bar{B}_\mu\right).
\label{f2}
\ee
\end{mathletters}
where
\be
\hat{B}_\mu^\prime(x)=
\bar{B}_\mu(x)-\epsilon_{\mu\nu\sigma\tau}
\int\dxp\partial^{\nu}f^{\sigma}\left(x-\xp\right)
\phi^{\tau}(\xp)
\ee
Substituting Eqs.~(\ref{f1}), (\ref{f2}) back into (\ref{loops})
\be
Z(\J)&=&\exp\Bigg\{\frac{i}{2}
\int(dx)(dx')\Bigg(J^\mu(x)D_{\mu\nu}\left(x-x^\prime\right)J^\nu\left(x^\prime\right)
+ {\Jg}^{\mu}\left(x\right)D_{\mu\nu}\left(x-x^\prime\right){\Jg}^{\nu}\left(x^\prime\right)\Bigg)
\nn
&& \hspace{2cm} \mbox{}
-i\epsilon_{\mu\nu\sigma\tau}\int(dx)(dx')(dx'')J_\kappa(x)
D^{\kappa\mu}\left(x-\xp\right)\partial^{\prime\nu}
f^\sigma\left(\xp-\xpp\right){\Jg}^\tau(\xpp)\Bigg\}
\nn
&& \hspace{-.5cm}
\times \int\dphi\dphip\exp\Bigg\{ i\int(dx)\Bigg(\phi_{\mu}(x)
\bigg[\frac{\delta}{i\delta \bar{A}_{\mu}(x)}
+\epsilon^{\mu\nu\sigma\tau}\int\dxp\partial_\nu 
f_\sigma\left(x-\xp\right)\frac{\delta}{i\delta\bar{B}^\tau(\xp)}\bigg]
+\phi^{\prime}_{\mu}(x)\frac{\delta}{i\delta\bar{B}_\mu(x)}\Bigg)
\nn
&&
\hspace{1cm}\mbox{}
- \frac{i}{2}\int(dx)(dx')
\Bigg(\phi^\mu(x)K_{\mu\nu}(x-\xp)\phi^\nu(\xp)-\phi^{\prime\mu}(x)
\tilde{\Delta}^{-1}_{\mu\nu}(x-\xp)\phi^{\prime\nu}(\xp)\Bigg)\Bigg\}
 F_1(\bar{A})F_2(\bar{B})\, ,
\ee
and performing the resulting quadratic integration over 
$\phi(x)$ and $\phi^\prime(x)$, we obtain the results in Eq.~(\ref{master}).

\end{document}